\documentstyle[12pt,epsfig]{article}
\newcommand{\XA}{x_{A}}
\newcommand{\xr}{x_{R}}
\newcommand{\xf}{x_{F}}
\newcommand{\xt}{x_{T}}
\newcommand{\xs}{x_{S}}
\newcommand{\ps}{p_{S}}
\newcommand{\PT}{p_{T}}
\newcommand{\AN}{A_{N}}
\newcommand{\Aq}{A_{q}}
\newcommand{\Pq}{P_{q}}
\newcommand{\eb}{E^{BEAM}}
\newcommand{\FPT}{F}
\newcommand{\GXA}{G}
\newcommand{\IA}{{h}_{1}}
\newcommand{\IB}{{h}_{2}}
\newcommand{\IC}{{h}_{3}}
\newcommand{\KS}{K^{0}_{S}}
\newcommand{\im}{Im}
\newcommand{\SNF}{snf}
\newcommand{\SF}{sf}
\begin{document}
\begin{flushright}
 Preprint IHEP-98-84\\
 hep-ph/0110152
\end{flushright}
\vskip 1cm
\centerline{ \bf\Large {A New Scaling for Single-Spin Asymmetry}}
\centerline{ \bf\Large {in Meson and Baryon Hadroproduction}}
\begin{center} V.V.Abramov$^{1}$ 
\end{center}
\vskip 0.5cm
\centerline { Experimental Physics Department,}
\centerline { Institute for High Energy Physics, P.O. Box 35, }
\centerline{ Protvino, 142281 Moscow region, Russia}
\centerline {$^{1}$E-mail: abramov\_v@mx.ihep.su}
\begin{abstract}
    Experimental data on analyzing power for inclusive meson and
 baryon production in hadron-proton(polarized) collisions have been analyzed.
 It is found that the existing data can be described  by a simple function
 of  collision energy ($\sqrt{s}$), transverse momentum ($\PT{}$)  and  a new 
 scaling  variable $\XA{}$ = $E/\eb{}$.  At beam energies above 40 GeV and
 $\PT{}$ above 1 GeV/c the analyzing power is described by a 
 function of $\XA{}$ and $\PT{}$ only ($\AN{}={\FPT}(\PT{}){\GXA}(\XA{})$)
 for both polarized proton fragmentation and central regions of proton-hadron 
 collision. Comparison of data from Fermilab and new IHEP data measured
 using 40 GeV/c polarized proton beam was most decisive for the revelation
 of the above regularities.  This new scaling law allows one to predict  
 the analyzing powers for kinematic regions, not yet explored  in  
 experiments and constrains models of strong interactions.  
 The new scaling law  allows  one also to use some reactions as
 polarimeters  for experiments with a polarized beam. 
\end{abstract}



\vskip 1.0cm
\centerline{Submitted to {\em Eur. Phys. J. C\/} }
%
\newpage
\section{Introduction}
  In this paper we will study from empirical point of view the existing world
data for one measured spin-dependent quantity (analyzing power) in
collisions of unpolarized hadrons with polarized protons and antiprotons.
The analyzing power ($A_{N}$), which is often called single-spin asymmetry,
should be distinguished from a raw asymmetry ($A_{RAW}$), which is directly
 measured in experiments and depends on a beam (or target) polarization 
$P_{B}$ ($P_{T}$) and a dilution factor $f$. For polarized beam experiments 
$A_{RAW} = A_{N}\cdot P_{B}$, and for polarized target experiments 
$A_{RAW} = A_{N}\cdot P_{T}/f$.

  Practically all existing data (with $p \ge 6$ GeV/c) at intermediate and
 high energies are used for the analysis. Comparison of the Fermilab data
\cite{PLB264}, measured at 200 GeV/c with new 40 GeV/c
 IHEP polarized beam data \cite{FODS} was an important step in the revelation
 of scaling features of the analyzing power.

 The data measured with meson beams using polarized targets 
\cite{Amag,APOK49,PLB243} are just briefly explored here.
 The important investigations in this field were done at the IHEP and other
 accelerators and merit probably a dedicated paper.  

  Recent measurements have shown that at high enough energies the
analyzing power for inclusive production of hadrons in reactions

 $$  h_{1}^{\uparrow} \! h_{2}\rightarrow h_{3} + X $$

where $h_{1}$, $h_{2}$ and $h_{3}$ are hadrons,
 is large and described by a simple function of kinematic variables and
 shows an approximate scaling
 in $x_{F} =2p^{*}_{Z}/\sqrt{s}$ for fragmentation region
of vertically polarized protons  and a scaling in
 $\xt{}=2\PT{}/\sqrt{s}$ for 
central region \cite{PLB264,FODS,PRL77}--\cite{PRL64}. It is larger in
 the fragmentation region of polarized protons (antiprotons) then 
in the central region.  Some authors have assumed, that 
for the analyzing powers a radial scaling takes place
($\xr{}=2p^{*}/\sqrt{s}$) \cite{PRL64,BE9751}. However, as will be shown below,
this assumption has not been confirmed. The purpose of this
study is to find a suitable scaling variable, that allows one to describe in a
 unified way the dependence of analyzing powers on kinematic variables 
in a wide  range of beam energies, transverse momenta, and angles of particle 
production.

 A thorough study of the existing data has shown that the analyzing power
for the inclusive $\pi^{+}$-meson production in $p^{\uparrow} \!p$ collisions 
has the following features \cite{PLB264,FODS,PRL64}:
\begin{itemize}
\item[\bf {a)}] the scaling and linear dependence on  $\xf{}$ or $\xt{}$ 
in the region  of polarized proton fragmentation or in the central region, 
respectively;
\item[\bf {b)}] the analyzing power maximum in the fragmentation region 
(near $\xf{}$=1) is  approximately two times higher than it is in the central 
region (near $\xt{}$=1);
\item[\bf {c)}] the analyzing power changes its sign (or is zero)
 in the polarized proton 
fragmentation  region at $\xf{}$ near  0.18,  whereas in the central region 
it takes place at $\xt{}$ near 0.37, which is approximately  two times  higher;
\item[\bf {d)}]  the analyzing power grows with $\PT{}$ rise at fixed
 $\xf{}$, has a plateau above 1 GeV/c, and  probably decreases when $\PT{}$
 gets much higher 1 GeV/c;
\item[\bf {e)}] the analyzing power is zero at $p_T$ = 0 due to the azimuthal
symmetry of cross section.
\end{itemize}
   Feature (d) has not too much experimental conformation yet, but below
it is assumed to be valid.
 
    The  features (a), (b) and  (c)  are well explained  if we assume 
that at high enough energy and $p_{T}$
 the analyzing power is described by a function of $\PT{}$ and
 a new scaling variable ($x_{A}$):
\begin{equation}
           \AN{}={\FPT}(\PT{}){\GXA}(\XA{}). \quad     \label{eq:FG}
\end{equation}
The scaling variable $x_{A}$ is defined as 
\begin{equation}
          \XA{} = E/\eb{}, \quad     \label{eq:XA}
\end{equation}
where $E$ and $\eb{}$ are
energies of the detected particle ($\pi^{+}$) and the beam particle 
(proton), respectively, in the 
laboratory frame, and a polarized beam particle  
collides with a  target at rest. 
This occurs because in the fragmentation region $\XA{}$ is close to $\xf{}$
and  its maximum is equal to 1.0, whereas in the central region  $\XA{}$
 is close to $0.5\cdot \xt{}$ and its  maximum is equal to 
0.5, when beam energy is divided between two high $\xt{}$ jets (particles).
 In case  of experiments
with a polarized target \cite{Amag,APOK49,PLB243,BE9751,PL94B,NPB142},
 $\XA{}$ is calculated in  anti-laboratory frame, where a beam
particle is again a transversely polarized proton.  
Eq.~(\ref{eq:XA}) takes the form 
$\XA{}=p_{\IC{}}\cdot p_{\IB}/p_{\IA}\cdot p_{\IB}$ when it 
is expressed in the Lorentz--invariant way.

  The eq.~(\ref{eq:FG}) means not only a scaling law for $A_{N}$, but in
addition a factorization of $p_{T}$ and $x_{A}$ dependences. 
This factorization simplifies the analysis and is in agreement with the
existing data, as will be shown below.

    We expect that most (but not all) of the specified above analyzing power
 features
(a -- e) are valid not only  for $\pi^+$ production, but also for other 
pseudoscalar mesons ($\pi^-$, $\pi^0$, $K^{\pm}$, $K_{S}$, $\eta$), as well as
for some baryons (protons, antiproton, hyperons), though the experimental 
information  for some of them is very limited. In particular, feature (e) is
 valid for any considered reaction, since the normal vector to the scattering
 plane is undefined when $p_{T}=0$, and no left-right asymmetry exists.
 Of course, at $x_{A}=0$ analyzing power is also zero, but this is not
 an independent feature, since in this case  $p_{T}=0$. 
 Feature (e) means that  ${\FPT}(0)=0$, but it does not meant that 
${\GXA}(0)=0$. In  particular, $A_{N}$ as a function of $x_{A}$ at fixed
value of $p_{T} \ne 0$ will not tends to zero when  $x_{A}$ approaches zero.
On the other hand if we consider $A_{N}$ measurements at fixed laboratory
angle, as often happens, $p_{T}  \propto \XA{}$ and  $A_{N}$ tends to zero
when  $x_{A}$ approaches zero.

     There are several alternative variables which are numerically  close 
to the $\XA{}$ variable, given by eq.~(\ref{eq:XA}). In particular,  
\begin{equation}
  \XA{}^{'}     = (\xf{}+\xr{})/2,       \label{eq:XFR}
\end{equation}
\begin{equation}
  \XA{}^{''}   =  (E + P_{Z})/
(\eb{}+ P^{BEAM}_{Z}),  
\label{eq:XPZ}
\end{equation}
\begin{equation}
  \XA{}^{'''} =  P/P^{BEAM},      \label{eq:XP}
\end{equation}
where $P$ and $P^{BEAM}$ are momenta  of the detected particle and 
beam particle, respectively, in the laboratory frame.
All of them are very close to each other at high energies and the
choice of the best scaling variable requires additional and very accurate 
measurements of  the analyzing power and kinematic variables. 
Eq.~(\ref{eq:XFR}) gives a very  transparent  explanation  of
 the $\xf{}$-scaling in the fragmentation region
 and the  $\xt{}$-scaling in the central region.

  The proposed scaling may be applied to the inclusive production of hadrons
in the collisions of polarized protons with light nuclei. Analyzing powers
measured in reactions $p^{\uparrow}p \rightarrow h + X$ and 
 $p^{\uparrow}d \rightarrow h + X$ , where $h$ is a charged hadron 
($\pi^{\pm}, K^{\pm}$, or $p$) agree within the errors \cite{PRD18}. Reactions
with pion beam and polarized proton or deuteron targets also give analyzing
power for $\pi^{0}$ production independent of the target within the 
errors \cite{PLB243}.

  Similar methods of different empirical scalings were used for the
 description of features of other reactions or observables.
  An example may be a description of the  analyzing power in $p^{\uparrow} C$
collisions with one outgoing charged particle. This reaction was often used
for the polarimetry purposes (see e.g.   \cite{pC} and references therein).
 In this and  other similar cases an empirical description of one of 
observables seem to be a correct way to show common characteristics as well
 as possible hidden features of strong interaction.

 A thorough study of  the available experimental data on
the analyzing powers is presented in the subsequent sections.
\section{ Analyzing power  for $p^{\uparrow} \!p\rightarrow \pi^{+} + X$
 reaction}
    For the study of scaling features of the analyzing power
all the available experimental data are presented in the frame in
which a polarized proton is a projectile with  spin directed upward
 and the target is at rest. 
The analyzing power is considered positive when more hadrons are produced
 to the left
in the horizontal plane looking  in the direction of the incident beam. Thus, 
 the original sign of the analyzing power for  experiments
 \cite{PL94B,NPB142}
 has been changed to the opposite one, in agreement with 
the definition given above. Kinematic variables for the experiments which
used polarized target have been transformed into the anti-laboratory frame. 
Unfortunately, not all authors   in their publications presented
a complete  set of variables ($\sqrt{s}$, $\PT{}$, $\xf{}$)
 for each point. For some 
experiments only limits on these variables are given that makes
transformation to other variables biased and limits accuracy of the 
$\XA{}$-scaling  check. Additional error ($\epsilon=\pm 0.025$) is added in 
quadrature  to all errors of $\AN$-values to take into account possible 
variable bias and  systematic errors during the fitting procedure below 
for $\pi^{+}$-meson production and other reactions if not stated otherwise.

\begin{figure}[ht]
\centerline{\epsfig{file=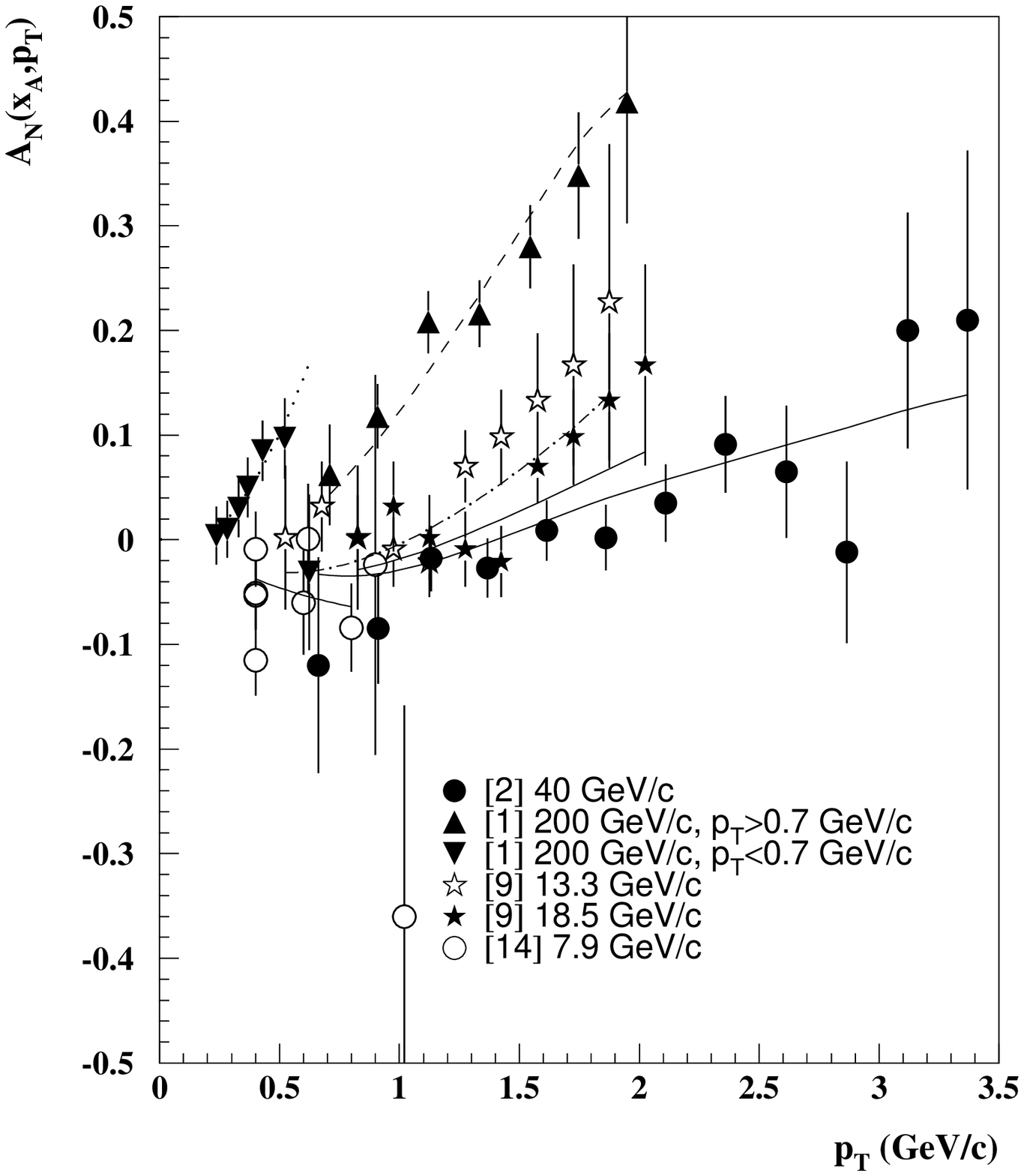,width=14cm}}
\caption {$\AN{}$ vs $\PT{}$ for the $\pi^{+}$ production by polarized protons.
      The curves correspond to a fit by eqs.~(6--10)  with 
the parameters given in Table~1.}
    \label{pi1pp_1pt}
\end{figure}

\begin{figure}[ht]
\centerline{\epsfig{file=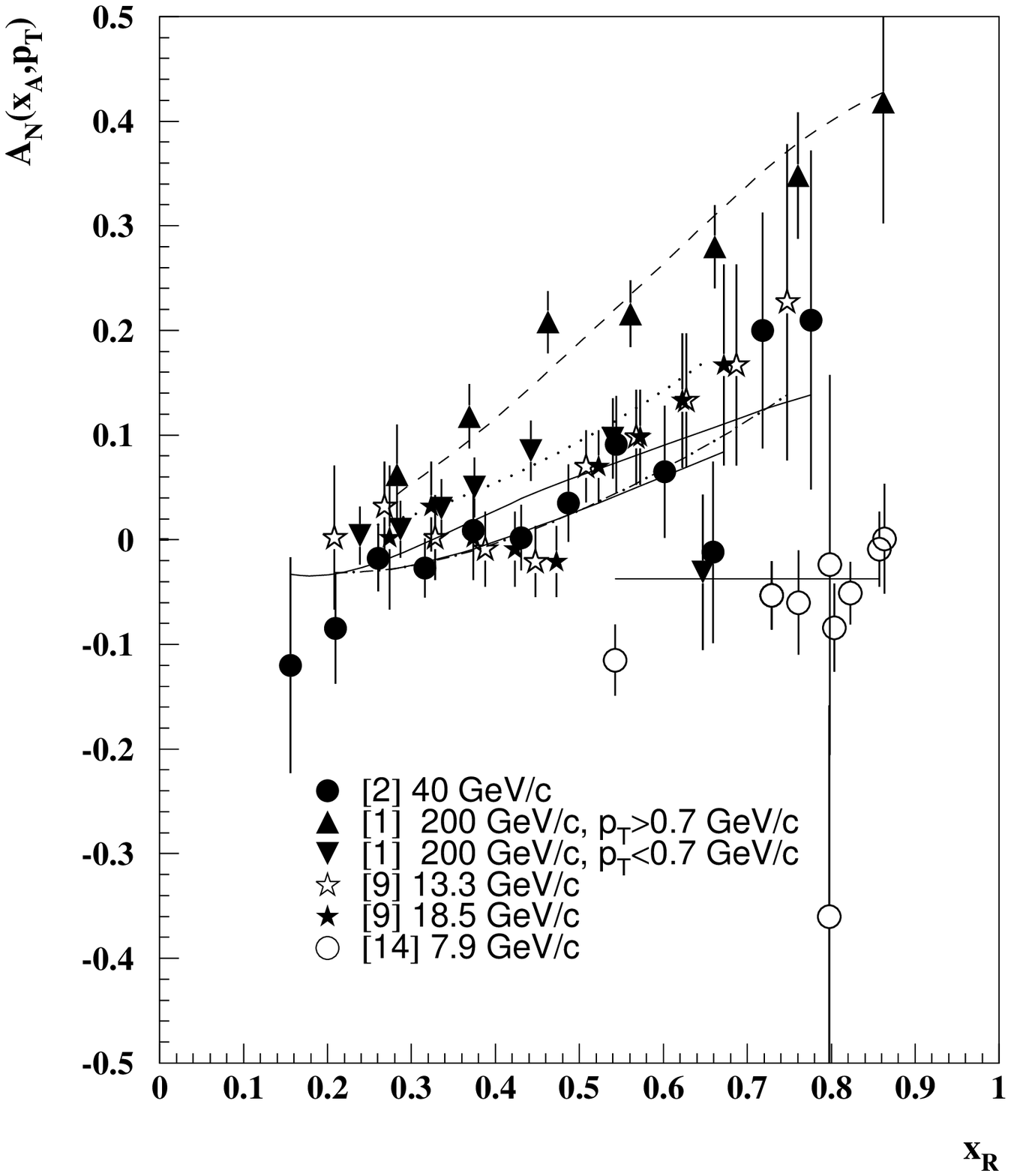,width=14cm}}
\caption {$\AN{}$ vs $\xr{}$ for the $\pi^{+}$ production by polarized protons.
      The curves correspond to a fit by eqs.~(6--10)  with 
the parameters given in Table~1.}
    \label{pi1pp_2xr}
\end{figure}

\begin{figure}[ht]
\centerline{\epsfig{file=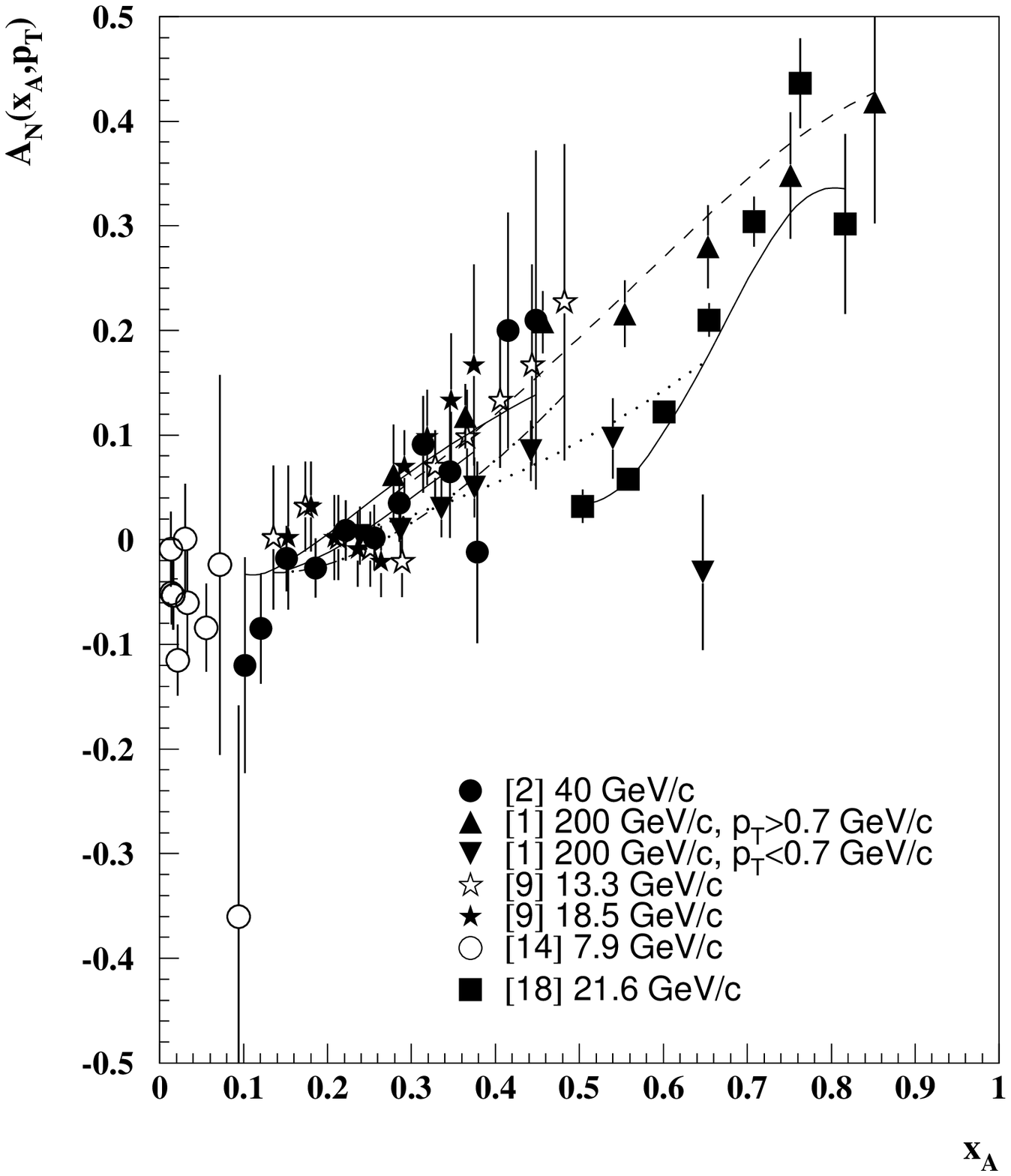,width=14cm}}
\caption {$\AN{}$ vs $\XA{}$ for the $\pi^{+}$ production by polarized protons.
      The curves correspond to a fit by eqs.~(6--10)  with
 the parameters given in Table~1.}
    \label{pi1pp_3xa}
\end{figure}

 The analyzing power of $\pi^{+}$ production in $p^{\uparrow} \!p$ 
collisions \cite{PLB264,FODS,PRL64,NPB142} 
is shown in Figs.~\ref{pi1pp_1pt}, \ref{pi1pp_2xr}  and \ref{pi1pp_3xa} as a
function of $\PT{}$, $\xr{}$,  and $\XA{}$, respectively. 
%
%
The highest $\PT{}$ ($\sim$3.5 GeV/c) is reached in \cite{FODS},
and the highest energy ($\sqrt{s}=19.43$ GeV) in \cite{PLB264}.
 As is seen in Figs.~\ref{pi1pp_1pt} and \ref{pi1pp_2xr},  
there is no scaling behaviour of the analyzing power as a function of
$\PT{}$ or $\xr{}$. Experiments, performed in forward, central and backward
regions have an analyzing power, decreasing from the forward to backward
 region, with the central region in the middle. In Fig.~\ref{pi1pp_3xa}
 the  analyzing power,
 as  a function of $\XA{}$, shows approximate scaling behaviour for all 
three regions, mentioned above. Only the subset of data \cite{PLB264} 
with $\PT{} < 0.7$ GeV/c
is below general trend, in agreement with the feature (d)  above.
 The analyzing power
dependence on $\XA{}$ is close to a linear one  in the consent with
the feature (a) above. 
A simple expression, which takes into account  all
the features (a--e) and low energy corrections can be used to fit
 the data shown in Fig.~\ref{pi1pp_3xa}:
\begin{equation}
  A_{N1}  = {\FPT}(\PT{})\cdot\cases 
 {  c\cdot\sin(\omega(\XA{} - x_{0})) +  
 a_{6}/s  , & if $\XA{} \ge a_{4}$;\cr
   c\cdot\sin(\omega((a_{4}-x_{0})+a_{5}(\XA{}-a_{4}))) +
 a_{6}/s  , & otherwise;\cr} \quad   \label{eq:AN1}
\end{equation}
where $x_{0}$ is a constant. 
The perturbative QCD predicts the vanishing of the analyzing power  at high
 $\PT{}$ \cite{RYSK,KANE}.
The same asymptotic has function  ${\FPT}(\PT{})$, which takes into account
the above mentioned features (d) and (e)
\begin{equation}
    {\FPT}(\PT{})  =  2{\PT}m/(m^{2} + \PT{}^{2}), \quad \label{eq:F1}
\end{equation}
where $\PT{}$ is measured in GeV/c
and $c,x_{0},m,a_{4} - a_{6}$ are free fit parameters. The exact shape of 
${\FPT}(\PT{})$ should be measured in future experiments.
 Parameters $a_{4},a_{5}$ and $a_{6}$ are equal to zero, and $\omega=1$ for 
$\pi^{+}$-meson production.  They are introduced for other reactions,
considered below, to take into account possible nonlinearity and 
non-asymptotic contribution  to the analyzing power at low energy. 
So, for $\pi^{+}$ at high enough energy we have 
${\GXA}(\XA{}) = c\cdot\sin(\omega(\XA{} - x_{0}))$.

The point $\XA{}=x_{0}$ 
may be interpreted as a point where the relative phase of two helicity 
amplitudes  (spin-flip and spin-nonflip) passes through zero and, perhaps, 
changes its sign, as was suggested in \cite{PLB243}. This problem will be
discussed in section 9. From experimental point of view the zero-crossing point
of the analyzing power was observed not only in the reaction of $\pi^{0}$
 production by $\pi^{-}$ beam~\cite{Amag,APOK49,PLB243}, but similar 
indications 
 were observed in some reactions of meson and baryon production by polarized
 proton beam~\cite{PLB264,FODS,PRD53,PRL64,PRD18,AD9756}. Experimental study
 of zero-crossing point is difficult because of small value of $A_{N}$ and low
setup efficiency near that point. The existence of  zero-crossing point
(with possible change of $A_{N}$ sign near it) may be critical for many
 theoretical models.

\begin{figure}[ht]
\centerline{\epsfig{file=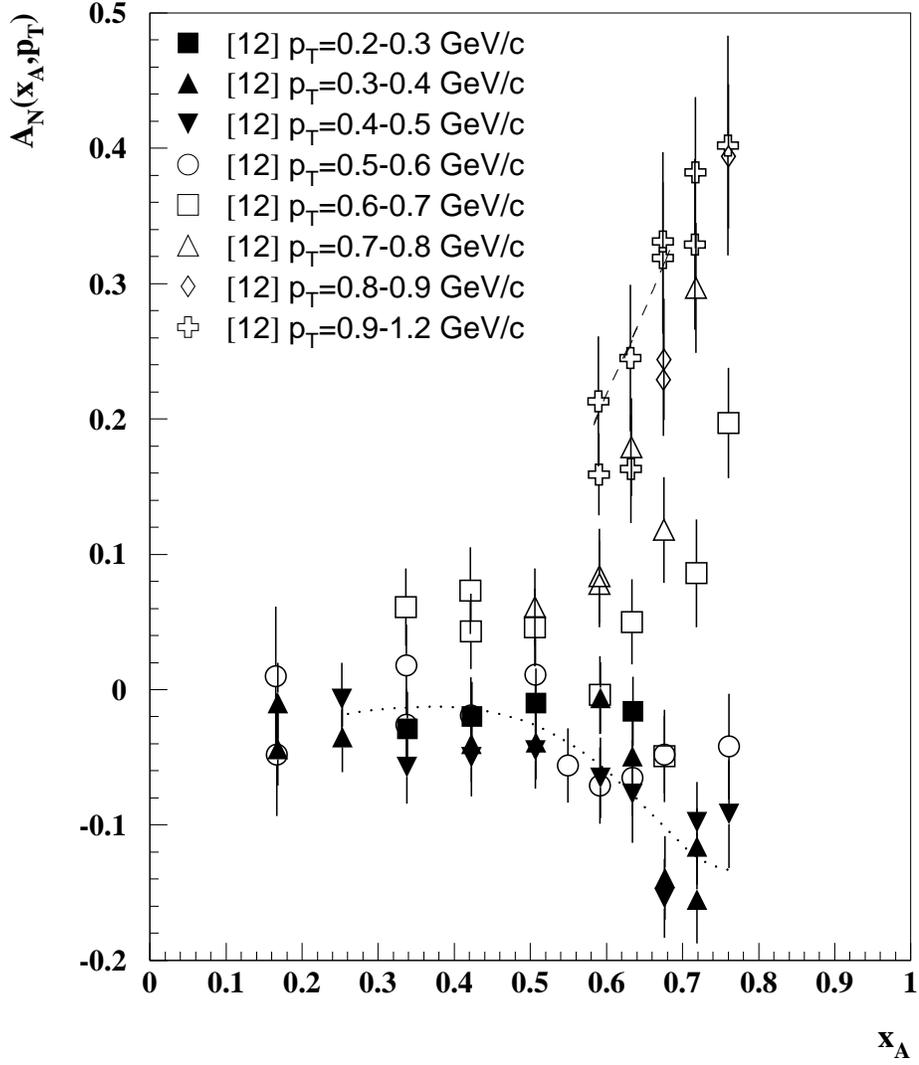,width=14cm}}
\caption {$\AN{}$ vs $\XA{}$ for the $\pi^{+}$ production  by polarized
 11.75 GeV/c protons [12]. Dotted and dashed curves correspond 
to a fit by eqs.~(6--10) 
for the regions $0.4 \le \PT{} \le 0.5$ and $0.9 \le \PT{} \le 1.2$ GeV/c,
respectively.}
    \label{pi1DR_4xa}
\end{figure}

\begin{figure}[ht]
\centerline{\epsfig{file=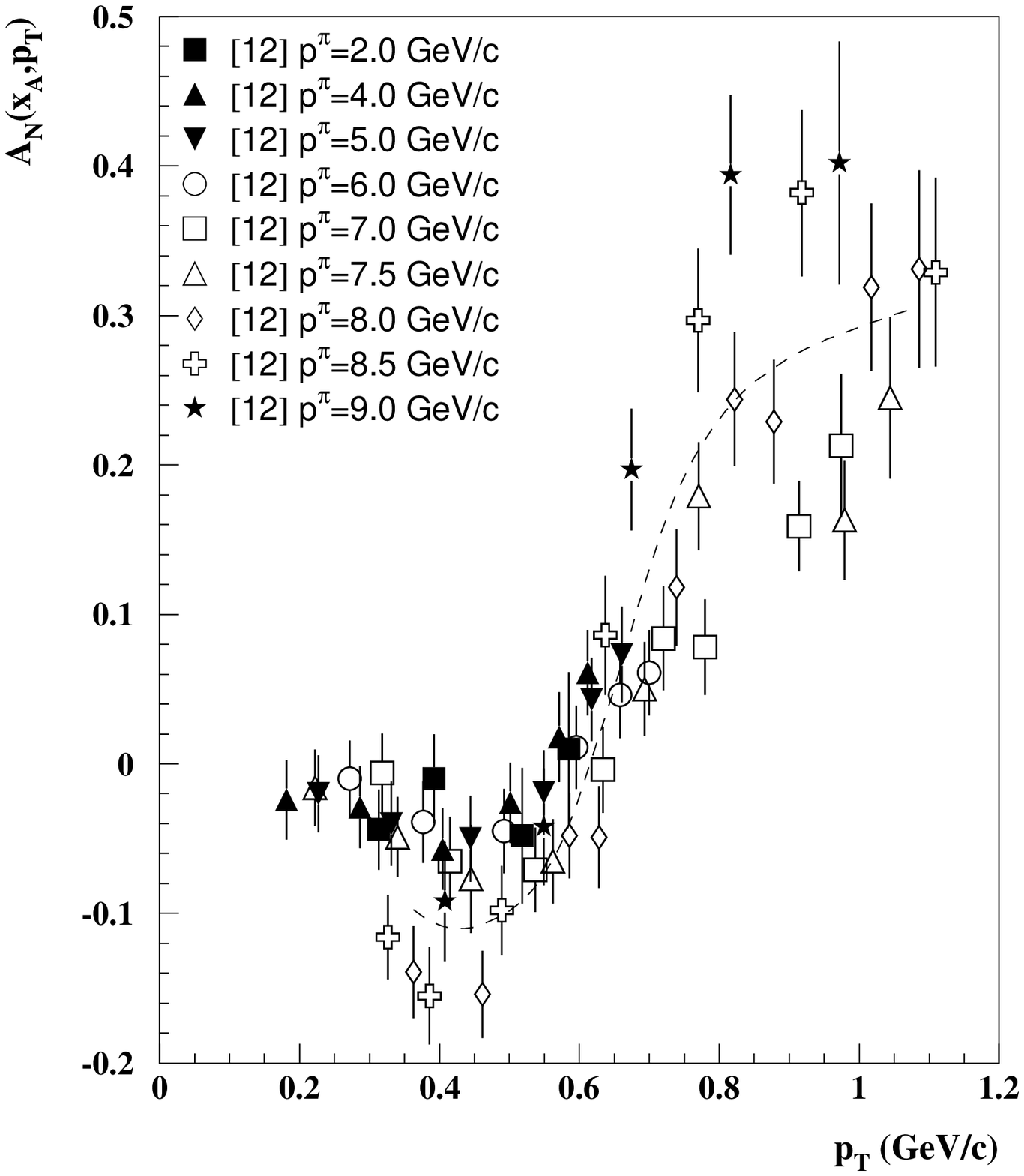,width=14cm}}
\caption {$\AN{}$ vs $\PT{}$ for the $\pi^{+}$ production  by polarized 
11.75 GeV/c protons [12]. The curve corresponds to a fit by eqs.~(6--10) 
for the $p^{\pi}=8$ GeV/c.}
    \label{pi1DR_5pt}
\end{figure}

     Along with the experiments  presented 
in Figs.~\ref{pi1pp_1pt}--\ref{pi1pp_3xa},  there is an experiment with very 
thorough measurements of the analyzing power   at 11.75 GeV/c \cite{PRD18}.
The measurements have been performed  for a set of  fixed secondary momenta,
corresponding to fixed $\XA{}$ values, and for each $\XA{}$ as a function
 of the production angle  or $\PT{}$.
The data are presented  in Figs.~\ref{pi1DR_4xa} and \ref{pi1DR_5pt}, 
as a function of   $\XA{}$ and $\PT{}$, respectively.
%
As is seen from Figs.~\ref{pi1DR_4xa} and \ref{pi1DR_5pt}, 
only the points corresponding to  the highest available $\PT{}$,
which are about 1 GeV/c,  are close to the scaling function (\ref{eq:AN1}) 
and to the experimental points
shown in Fig.~\ref{pi1pp_3xa} for higher energies. 
Dependence of $\AN{}$ on $\PT{}$ is  very different from the corresponding
behaviour at higher energies, shown in Fig.~\ref{pi1pp_1pt}. 
To understand this  difference of data \cite{PRD18} from the rest of the data,
we have to assume that at 11.75 GeV/c ($\sqrt{s}=4.898$ GeV) and  low $\PT{}$ 
there  exists an additional contribution to the analyzing power,
which is approximated by the expression
\begin{equation}
  A_{N0} ={\FPT}_{0}(\PT{}) \Bigl( b_{1}\tanh( b_{2}(\PT{}-b_{7}) )
  \sin(b_{8}\XA{}^{{b}_{4}}) +
  b_{5} + b_{6}\XA{} \Bigr),    \label{eq:AN0}
\end{equation}
where function ${\FPT}_{0}(\PT{})$ suppresses the analyzing power at low
 $\PT{}$
\begin{equation}
    {\FPT}_{0}(\PT{})  = 2\PT{}^{2}/(b_{3}^{2} + \PT{}^{2}),
 \quad \label{eq:F0}
\quad
\end{equation}
and $b_{1}-b_{8}$ are free parameters.

Fit of a combined data set, which includes the data, presented in 
Figs.~\ref{pi1pp_3xa} and  \ref{pi1DR_4xa}, 
requires additional  assumption  that the $A_{N0}$ 
contribution decreases  with energy, and the complete  analyzing power is 
\begin{equation}
  \AN{}   =  A_{N1}  +  
   A_{N0}\cdot(4.898/\sqrt{s})^{{b}_{9}} , \label{eq:AN}
\end{equation}
where $b_{9}$ is a free parameter.

 The results of the combined data set fit are presented in 
Figs.~\ref{pi1pp_3xa}  and \ref{pi1DR_4xa}  (corresponding curves) and 
in Table~\ref{Tab1pi+}  (fit parameters). Two subsets  of 
the combined data  are shown in the separate figures to give a clearer
representation  of 117 data points. Parameter $\omega$ was fixed since the data
show a linear dependence on $\XA{}$ and the experimental accuracy is not 
sufficient to get  $c$ and $\omega$ values separately. In all the fits below
it is assumed that $\omega = 1$, unless otherwise specified.
\begin{table}[htb]
\small
\caption{Fit parameters of eqs.~(6)--(10) for $\pi^+$-mesons.}
\begin{center}
\begin{tabular}{ccccc}    \hline
 $c$          & $x_{0}$       & $m$    & $a_{6}$ \\[0.3cm] \hline
0.69 $\pm$0.08 &0.170$\pm$0.046&2.0  $\pm$0.4  &0.00         \\[0.3cm]\hline
$\omega$       &$b_{1}$         & $b_{2}$       & $b_{3}$    \\[0.3cm] \hline
1.00          &0.148$\pm$0.029 &8.6 $\pm$2.3 &0.35 $\pm$0.07 \\[0.3cm]\hline
   $b_{4}$    &$b_{5}$         & $b_{6}$       & $b_{7}$  \\[0.3cm] \hline
4.8 $\pm$1.0 &0.004 $\pm$0.015 &-0.148$\pm$0.041&
0.646$\pm$0.016\\[0.3cm]\hline
  $b_{8}$      &   $b_{9}$     & N points    & $\chi^{2}$ \\[0.3cm] \hline
  5.6 $\pm$2.6  & 2.0$\pm$1.9    &    117      & 114.4     \\[0.3cm]\hline
 \end{tabular}
\end{center}
\label{Tab1pi+}
\end{table}
The agreement between  the fitting curves and the data is rather  good.
The analysis has shown that the contribution of $A_{N0}$ term to 
(\ref{eq:AN}) is small ($ \le 0.08$) for the experiments  presented 
in Fig.~\ref{pi1pp_3xa}. On the other hand, the term $A_{N1}$ is
significant  ($\le 0.3$) for a kinematic region of the experiment 
 \cite{PRD18},  presented in Figs.~\ref{pi1DR_4xa} and \ref{pi1DR_5pt}.

\begin{figure}[ht]
\centerline{\epsfig{file=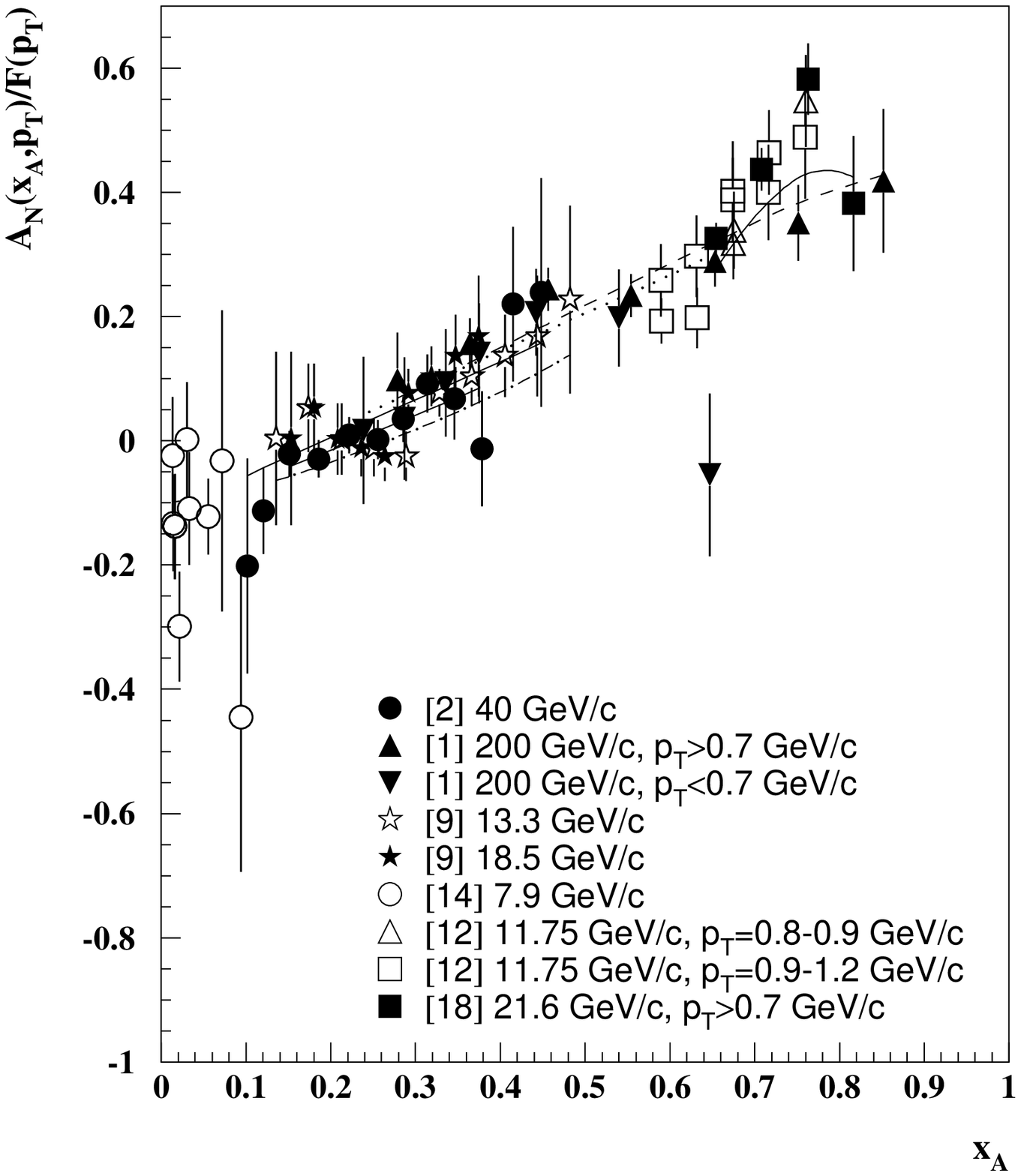,width=14cm}}
\caption {The ratio $\AN{}/{\FPT}(\PT{})$ vs $\XA{}$ for the $\pi^{+}$  
production  by polarized  protons. The curves correspond to a fit by 
eqs.~(6--10)  with  the parameters given in Table~1.}
    \label{pi1ra_6xa}
\end{figure}

The ratio of the experimental analyzing power and  ${\FPT}(\PT{})$, which is
 expected to be a function of $\XA{}$ only, with a possible small dependence
 on  $\sqrt{s}$, is shown in Fig.~\ref{pi1ra_6xa}. 
%
The data from \cite{PRD18} are presented in Fig.~\ref{pi1ra_6xa}
by two subsets, corresponding to $0.8\le \PT{} \le 0.9$ GeV/c and
$0.9\le \PT{} \le 1.2$ GeV/c, respectively.
 All the experimental points in Fig.~\ref{pi1ra_6xa}
 are  consistent with  the simple function of $\XA{}$
\begin{equation}
 \AN{}/{\FPT}(\PT{}) = c\cdot \sin(\omega(\XA{} - x_{0})), \label{eq:ANLIM}
\end{equation}
that confirms scaling behaviour and factorization of $\PT{}$ and $\XA{}$  
dependencies, assumed in (\ref{eq:FG}) and (\ref{eq:AN1}) at
 high $\PT{}$ and high beam energy.

 Recently, when the this paper was already prepared for publication,
 new 21.6 GeV/c data for  $\pi^{+}$, $\pi^{-}$ and proton production 
 analyzing powers in $p^{\uparrow} C$ collisions from the BNL E925 experiment
 have been measured \cite{E925}, which confirm the  $\AN{}$ behaviour, 
predicted by eqs. (6-10). In particular, the value of $\XA{}$, where  $\AN{}$
approaches to zero, is much higher due to non-asymptotic contribution
 (\ref{eq:AN0}) in low $p_{T} \le 0.7$ GeV/c region. Corresponding points are
shown in Figs. \ref{pi1pp_3xa} and \ref{pi1ra_6xa} along with predictions from
eqs. (6-10). The last four points with $p_{T} \ge 0.7$ GeV/c are compatible
 with general scaling behaviour of other data shown in Fig.~\ref{pi1ra_6xa}.
It has to be noted that only statistical errors are shown for data \cite{E925}.
The overall statistical and systematic error in the beam polarization gives a 
relative scale uncertainty of 24\% for $A_{N}$, the same for all  three
reactions of interest for all $x_{F}$ and $p_{T}$. Due to this scale
 uncertainty and the usage of different target (carbon) these data are not
 included in the overall fit and are shown for the purpose of comparison only.

   The results of the fit  (\ref{eq:AN}) show that  the
data sample  \cite{PRD18}  can be compatible with the rest of the data 
assuming that the additional contribution (\ref{eq:AN0}) is significant only
at low beam energy  and $\PT{}$. The physical nature of this contribution,
which is negative at $\PT{}$ near 0.4 GeV/c even at high $\XA{}$,  is not
completely  clear. It could be a resonance contribution \cite{RYSK,RESON},
or something  else. The authors of \cite{PRD18} have assumed that the observed
 analyzing power is explained by the baryon exchange in $u$-channel.

The existing experimental data at higher energies, presented  in  
Fig.~\ref{pi1pp_3xa}, are not very sensitive  to the contribution 
(\ref{eq:AN0}), which is prominent at 11.75 GeV/c. 
A detailed experimental study of region $\PT{} \le 1$ GeV/c at higher energies
and different production angles could help to understand its nature.

 Fit parameters of eq. (\ref{eq:AN1})  for
 different definitions of scaling variable
 (\ref{eq:XA}) -- (\ref{eq:XP}) are presented in Table~\ref{Tab2pi+}. 
Only parameters $c, x_{0}, m$ are free here. All other parameters 
are the same as in Table~\ref{Tab1pi+}.
The difference in  $\chi^{2}$ is not very significant,
with a weak preference for eqs.~(\ref{eq:XA}), (\ref{eq:XPZ}) 
and (\ref{eq:XP}) variables.
\begin{table}[htb]
\small
\caption{Fit parameters of eq.~(6) for $\pi^{+}$-mesons. 
Different definitions of the scaling variable $\XA{}$ are used 
for comparison (eqs.~(2)--(5)). }
\begin{center}
\begin{tabular}{ccccc}    \hline
eq.&  $c$        &      $x_{0}$     & $m$
 & $\chi^{2}$ \\[0.3cm] \hline
(2)& 0.69 $\pm$0.08  &  0.170$\pm$0.047 &2.0 $\pm$0.4 &
 114.4 \\[0.3cm] 
(3)& 0.74 $\pm$0.07  &  0.166$\pm$0.013 &2.2 $\pm$0.3 &
 120.4 \\[0.3cm] 
(4)& 0.69 $\pm$0.07  &  0.167$\pm$0.013 &2.1 $\pm$0.3 &
 114.6 \\[0.3cm] 
(5)& 0.68 $\pm$0.06  &  0.170$\pm$0.013 &2.0 $\pm$0.2 &
 114.2\\[0.3cm] \hline
 \end{tabular}
\end{center}
\label{Tab2pi+}
\end{table}

  The error ($\epsilon= \pm 0.025$), added in quadrature to the error of
 $A_{N}$ at each data
 point during the fitting procedure, has not changed the fit parameters
 significantly, but has reduced $\chi^{2}$ by about a factor of two up to a
 level of about  unity per degree  of freedom. Errors, shown in figures,
 representing  experimental data, also include this additional error.
\section{Analyzing power for $p^{\uparrow} \!p \rightarrow \pi^{-} + X$ 
reaction}
\begin{figure}[ht]
\centerline{\epsfig{file=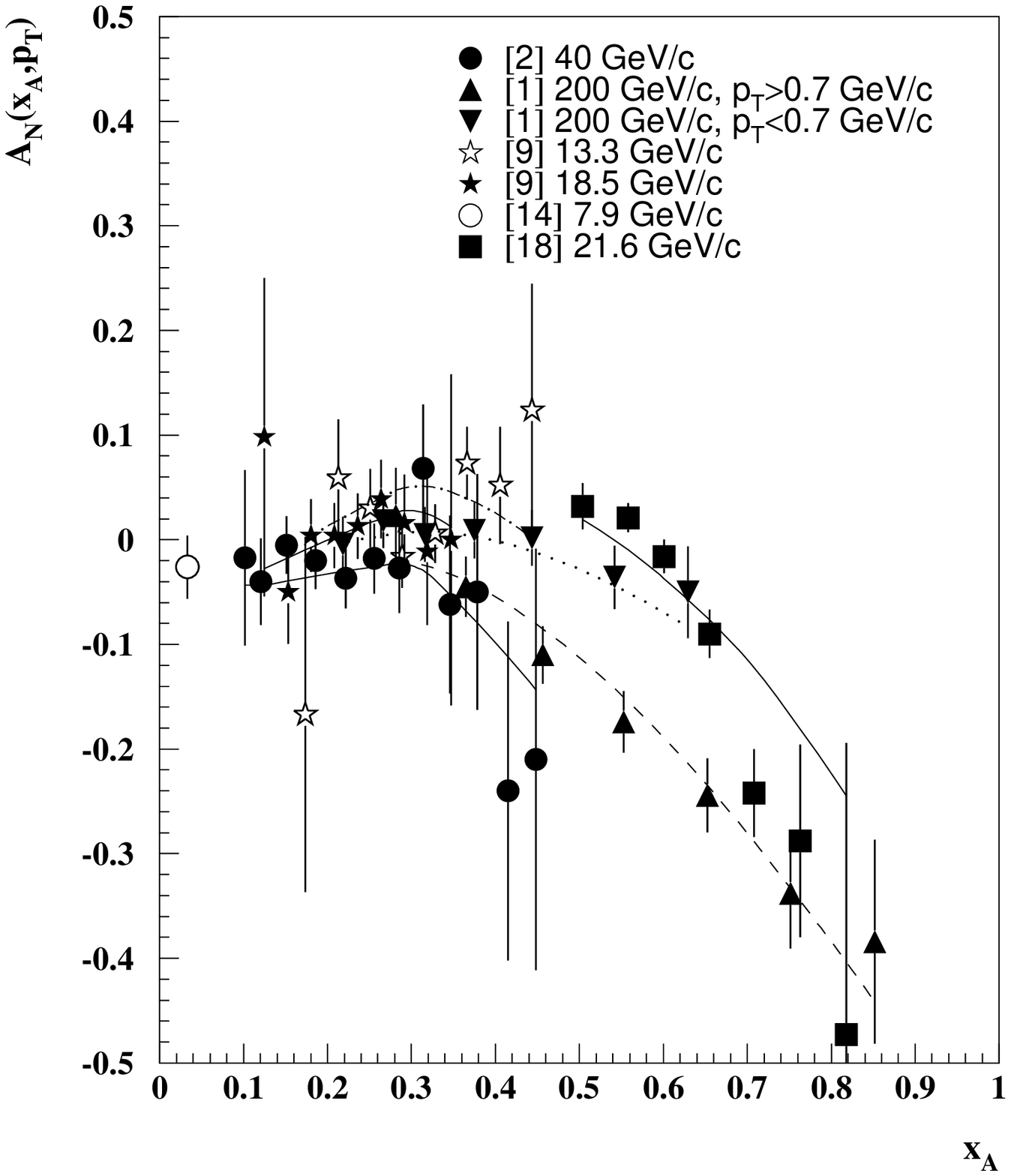,width=14cm}}
\caption {$\AN{}$ vs $\XA{}$ for the $\pi^{-}$ production by polarized protons.
      The curves correspond to a fit by eqs.~(6--10)  with the parameters
 given in Table~2.}
    \label{pi2pp_7xa}
\end{figure}

  The analyzing power for $\pi^{-}$-meson production by polarized protons 
 \cite{PLB264,FODS,PRL64,NPB142}
is shown in Fig.~\ref{pi2pp_7xa} as a function of $\XA{}$. 
As with $\pi^{+}$-mesons,
we observe an approximate scaling in the dependence of $\AN{}$ vs $\XA{}$. 
%
Selection of the data
with $\PT{} \ge 0.8$ GeV/c and $\eb{} \ge 40$ GeV leads to a good
agreement between two experiments \cite{PLB264,FODS} which implies their
scaling behaviour. 

 The new 21.6 GeV/c data for  $\pi^{-}$ production 
 analyzing power in $p^{\uparrow} C$ collisions from the BNL E925 experiment
 \cite{E925} are also shown in Fig.~\ref{pi2pp_7xa}  along with predictions 
from eqs. (6-10).  The last three  points with $p_{T} \ge 0.8$ GeV/c are
 compatible with general scaling behaviour observed at higher energies
 \cite{PLB264,FODS}. Low $p_{T} \le 0.8$ GeV/c points deviate from the scaling
law due to a non-asymptotic contribution (\ref{eq:AN0}). This is also a reason
 why $A_{N}$ cross zero level at much higher value of $x_{A} \approx 0.6$.
Only statistical errors are shown for data \cite{E925}, while overall
relative scale uncertainty for $A_{N}$  is 24\% .

\begin{figure}[ht]
\centerline{\epsfig{file=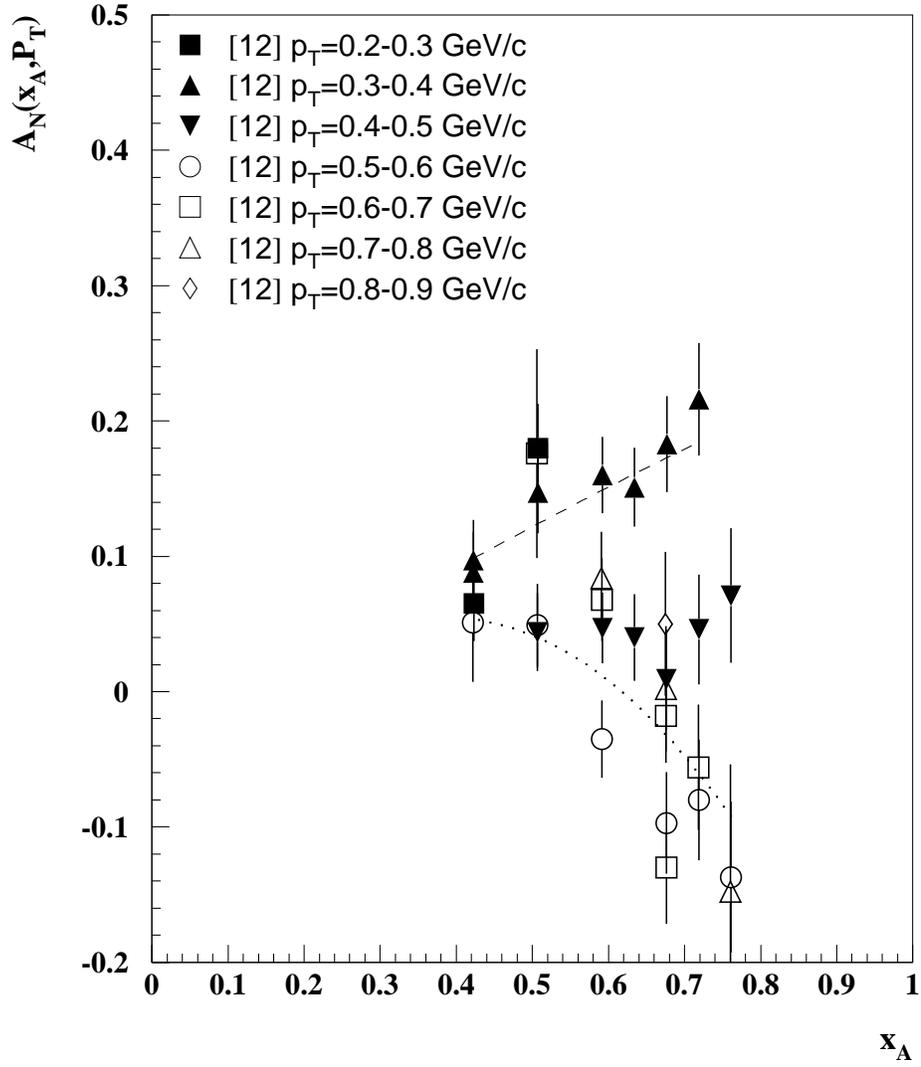,width=14cm}}
\caption {$\AN{}$ vs $\XA{}$ for the $\pi^{-}$ production by polarized 
11.75 GeV/c protons [12]. The dashed and dotted curves correspond to a fit by 
eq.~(6--10) for the regions 
$0.3 \le \PT{} \le 0.4$ and $0.5 \le \PT{} \le 0.6$ GeV/c, respectively.}
     \label{pi2DR_8xa}
\end{figure}

\begin{figure}[ht]
\centerline{\epsfig{file=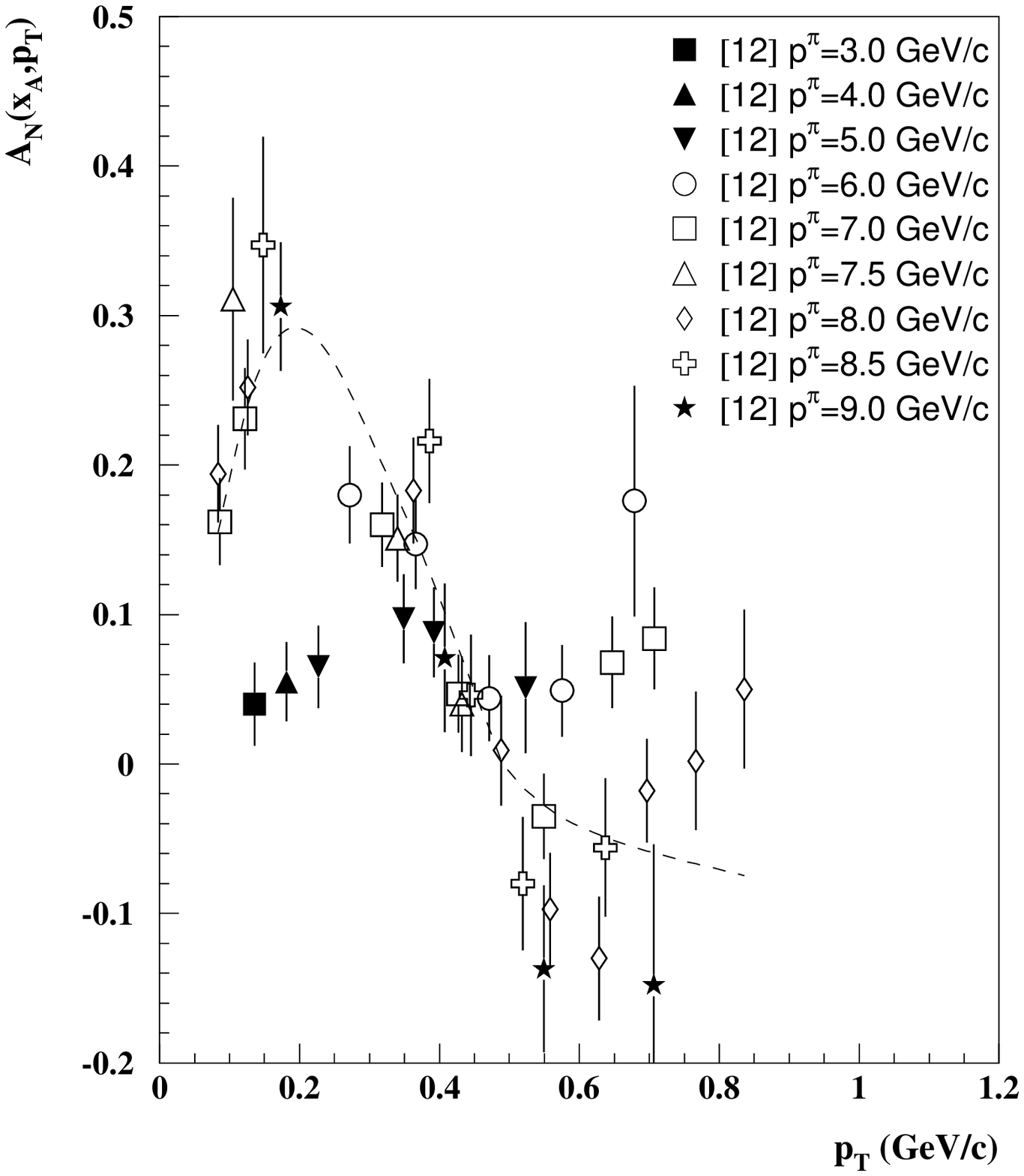,width=14cm}}
\caption {$\AN{}$ vs $\PT{}$ for the $\pi^{-}$ production by polarized 
11.75 GeV/c protons [12]. The curve corresponds to a fit by eqs.~(6--10) 
for the $p^{\pi}=8$ GeV/c.}
    \label{pi2DR_9pt}
\end{figure}

 Experiment  \cite{PRD18} reveals quite 
different $\XA{}$ and $\PT{}$-dependencies  at 11.75 GeV/c, 
in Figs.~\ref{pi2DR_8xa} and \ref{pi2DR_9pt}, respectively.
%
%
As with $\pi^{+}$, the greatest deviation from the scaling behaviour occurs at
low $\PT{}$. At $\PT{}=0.15$ GeV/c the analyzing power is very large and
 positive  in contrast to the large energy behaviour, where it is negative.
 One of possible origins of this  low energy analyzing power is probably
 the same as that 
discussed above for $\pi^{+}$-mesons, and its approximation is given by 
eqs.~(\ref{eq:AN1}) -- (\ref{eq:AN}).  The difference is that parameters 
$a_{4}$ and $a_{6}$ are now not equal to zero, while $a_{5}=0$.
The non-linear dependence of $A_{N}$ vs $x_{A}$ is taken into account
by setting $a_{4} > 0$ in eq. (\ref{eq:AN1}).
 Fit parameters of the combined data sample, shown in 
Figs.~\ref{pi2pp_7xa} and \ref{pi2DR_8xa}, are presented in 
Table~\ref{Tab1pi-}.
\begin{table}[htb]
\small
\caption{Fit parameters of eqs.~(6)--(10) for $\pi^-$-mesons.}
\begin{center}
\begin{tabular}{cccc}    \hline
 $c$          & $x_{0}$       & $m$    & $a_{4}$ \\[0.3cm] \hline
-0.96 $\pm$0.20 &0.185$\pm$0.075& 4.80   & 0.303$\pm$0.045 \\[0.3cm]\hline
 $a_{6}$          & $b_{1}$       & $b_{2}$    & $b_{3}$ \\[0.3cm] \hline
3.8  $\pm$1.8&-0.345 $\pm$0.089 &8.0 $\pm$2.8 &0.115$\pm$0.024\\[0.3cm]\hline
$b_{4}$         &   $b_{5}$    & $b_{6}$   & $b_{7}$    \\[0.3cm] \hline
3.1 $\pm$0.5 &-0.047 $\pm$0.018 &0.256$\pm$0.052&0.344$\pm$0.028\\[0.3cm]\hline
 $b_{8}$     &$b_{9}$    &N points          & $\chi^{2}$ \\[0.3cm] \hline
1.12$\pm$0.27  &0.76  $\pm$0.39   &    84       &   89.5     \\[0.3cm]\hline
 \end{tabular}
\end{center}
\label{Tab1pi-}
\end{table}
 Some of the parameters could not be well determined from the existing data
 and were fixed  ($m=4.8$, $\omega=1$) during the fitting procedure. 
The role of energy-dependent term 
($a_{6}/s$) is more significant for $\pi^{-}$, than for $\pi^{+}$ mesons.
Possible explanation can be related to resonance contribution \cite{RESON}.
The analyzing power in low $\XA{} \le 0.3$ region is close to zero in
 agreement with the expected large gluon contribution \cite{RYSK}.
\section{Analyzing power for $p^{\uparrow} \!p \rightarrow p + X$ reaction}
\begin{figure}[ht]
\centerline{\epsfig{file=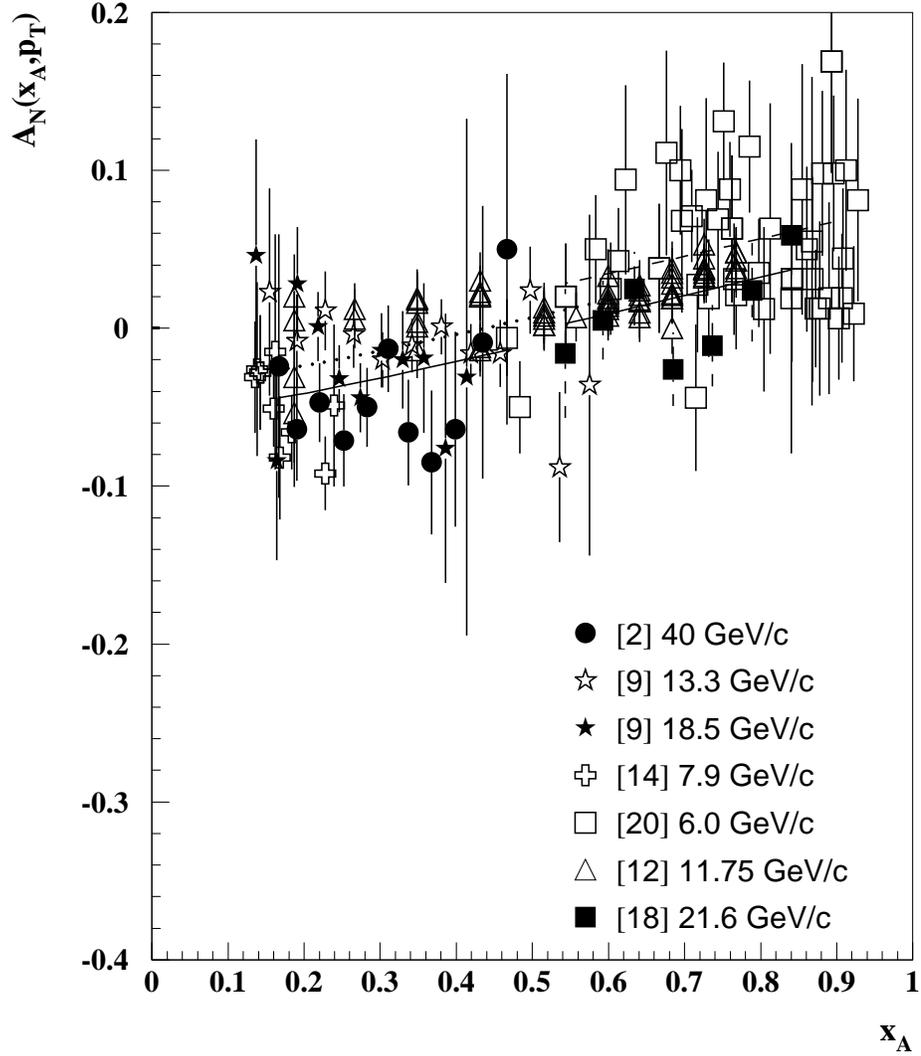,width=14cm}}
\caption {$\AN{}$ vs $\XA{}$ for the proton production 
 by polarized protons. The solid fitting curve corresponds to the 40 GeV/c 
data  [2]. The dotted curve corresponds to the 13.3 GeV/c data [9]. 
The dashed curve corresponds to the 6 GeV/c data [20]. The dash-dotted curve
corresponds to the 21.6 GeV/c data [18].}
   \label{p5pp_10xa}
\end{figure}

  The analyzing power for proton production has been measured at 6 different
 beam energies, from 6 up to 40 GeV \cite{FODS,PRL64,PRD18,NPB142,AYRES}.
 It is shown in Fig.~\ref{p5pp_10xa} as a function of $\XA{}$. The absolute 
value of $\AN{}$ is small ($\le 0.1$) and with the existing accuracy
$\AN{}$ is compatible with the approximate $\XA{}$-scaling, especially, when
taking into account possible systematic errors of the order of 0.02.
%
Nevertheless, the data fitting function (eq.~(\ref{eq:AN1})) is modified
to give a better approximation. In particular, the fit approximates the data 
better if a fitting function is not suppressed at high $\PT{}$, as is the case
with eq.~(\ref{eq:F1}). Non-asymptotic contribution to $\AN{}$ at low energies
is more significant for protons than for $\pi^{-}$-mesons and 
was approximated by  $a_{6}/s^{0.5}$ term. Eqs.~(\ref{eq:ANp})
 and (\ref{eq:F1p}) are used to fit the proton production analyzing power
\begin{equation}
  \AN{}  = {\FPT}_{P}(\PT{})( c\cdot\sin(\omega(\XA{} - x_{0})) 
+ a_{6}/s^{0.5} ), \quad    \label{eq:ANp}
\end{equation}
where 
\begin{equation}
    {\FPT}_{P}(\PT{})  =  1 - \exp (-\PT{}/m). \quad 
 \label{eq:F1p}
\end{equation}
Function ${\FPT}_{P}(\PT{})$ makes valid feature (e) of zero $A_{N}$
at $p_{T}=0$. 
  An extra error $\epsilon= \pm 0.015$ is added to the error of $A_{N}$
 at each data point.
The comparison of fit parameters  for different definitions of $\XA{}$,
given by eqs.~(\ref{eq:XA})--(\ref{eq:XP}), is shown in Table~\ref{PTAB1}.
\begin{table}[htb]
\small
\caption{Fit parameters of eqs.~(12--13) for the protons and different
definitions of the scaling variable $\XA{}$, eqs.~(2) --(6). Parameters 
$a_{4}$ -- $a_{5}$ are set equal to zero and $\omega=1$ during the fit.}
\begin{center}
\begin{tabular}{cccccc}    \hline
 eq. &  $c$        &      $x_{0}$     & $m$      & $a_{6}$
 & $\chi^{2}$ / points \\[0.3cm] \hline
 (2)  & 0.116$\pm$0.011 &  0.81 $\pm$0.15  &0.184 $\pm$0.006& 0.216$\pm$0.080
& 120.9/ 150\\[0.3cm] 
 (3)  & 0.117$\pm$0.012 &  0.90 $\pm$0.13  &0.186$\pm$0.007& 0.316$\pm$0.078
& 125.6/ 150 \\[0.3cm] 
 (4)    & 0.117$\pm$0.011 &  0.82 $\pm$0.14  &0.187$\pm$0.006&0.230$\pm$0.080
& 118.6/ 150 \\[0.3cm] 
 (5)   & 0.117$\pm$0.011 &  0.83 $\pm$0.14  &0.187 $\pm$0.006&0.236$\pm$0.079
& 119.4/ 150\\[0.3cm] \hline
 \end{tabular}
\end{center}
\label{PTAB1}
\end{table}
 The best $\chi^{2}$ is reached if $\XA{}$ is given by eq.~(\ref{eq:XPZ}).
 The analyzing power slightly rises with $\XA{}$ increase and changes its sign
near $\XA{}=0.5$ at beam energies around 10 GeV. Additional measurements of 
$\AN{}$ for protons at higher energies in the fragmentation region 
of polarized  protons could help to clarify a possible energy dependence
 of the analyzing power.

 The new 21.6 GeV/c data for  proton production 
 analyzing power in $p^{\uparrow} C$ collisions from the BNL E925 experiment
 \cite{E925} are also shown in  Fig.~\ref{p5pp_10xa}  along with predictions 
from eqs. (\ref{eq:ANp}-\ref{eq:F1p}).  The data are
 compatible with general trend of $A_{N}$ rise with increase of $x_{A}$.
Only statistical errors are shown for data \cite{E925}, while overall
relative scale uncertainty for $A_{N}$ is 24\%.

\section{Analyzing powers for $\pi^{0}$, $K^{+}$, $K^{-}$ and $\bar{p}$
 production by polarized protons}
\begin{figure}[ht]
\centerline{\epsfig{file=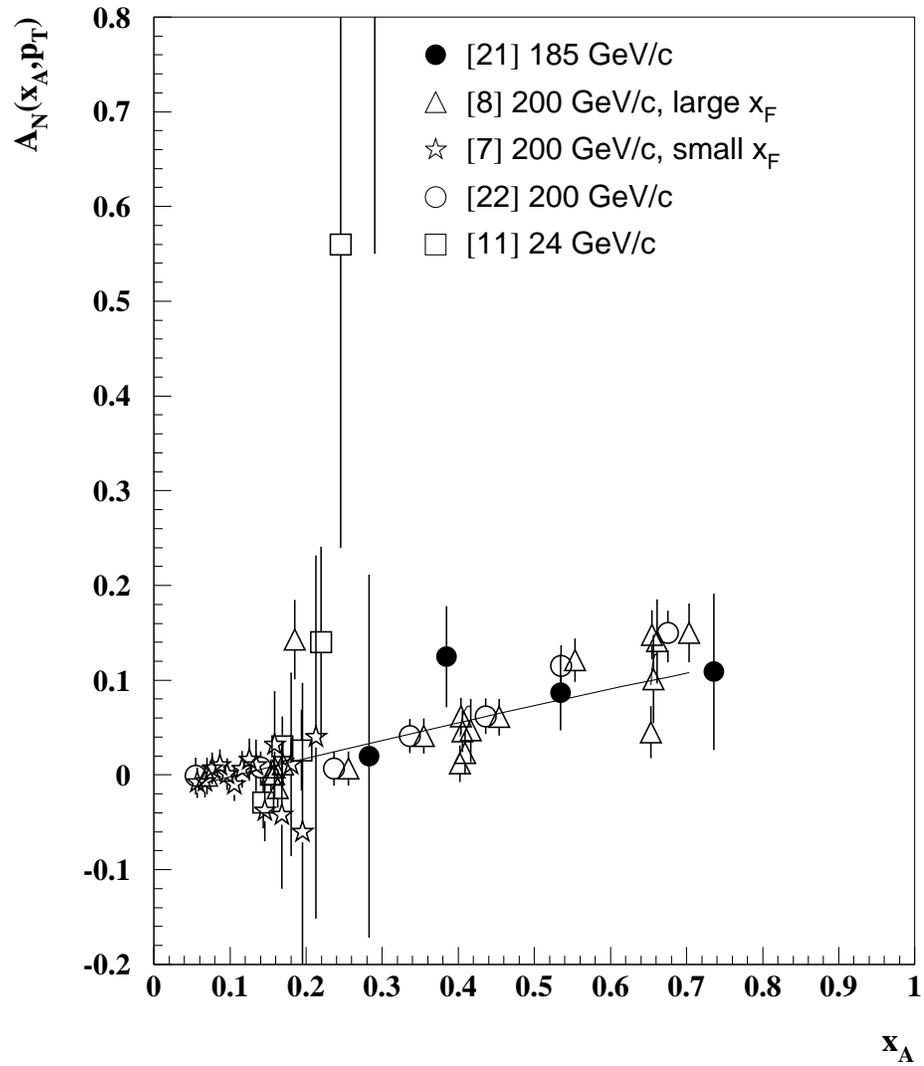,width=14cm}}
\caption {$\AN{}$ vs $\XA{}$ for the $\pi^{0}$ production by polarized 
protons. The fitting curve corresponds to the 200 GeV/c data [8].}
    \label{pi7pp_11xa}
\end{figure}

The analyzing power for $\pi^{0}$-meson production in $p^{\uparrow} \!p$
 collisions has been  measured at 24, 185 and 200 GeV/c 
\cite{PRD53,ZPC56,PL94B,PRL61,PL261}. 
The data are shown in Fig.~\ref{pi7pp_11xa}
as a function of $\XA{}$. They are compatible with a simple dependence
given by eq.~(\ref{eq:AN1}) with $a_{4}=0$ and $a_{6}=0$.
 The fit parameters are shown in Table~\ref{KTAB1}.
The data  \cite{PL94B} were measured using a polarized target, where
 the dilution factor plays an important role, reaches large values (and also
errors) and may be badly determined. 
A very large analyzing power observed in a few points with largest $p_{T}$
 at 24 GeV/c \cite{PL94B} probably  results from
the above problem of dilution factor measurement. 

 Assumption of
the $\XA{}$-scaling allows one to explain the enigma of the E704 data
\cite{PRD53}, which have not shown any significant analyzing power, though
 experiment has reached high $\PT{}$ values up to 4.5 GeV/c. This is because
the corresponding values of  $\XA{}$ are near $x_{0}=0.111$, where $\AN{}$
as a function of $\XA{}$ is close to zero. Both, the high $\PT{}$ \cite{PRD53},
 and the  high $\xf{}$ \cite{ZPC56} data are in good agreement if plotted
vs  $\XA{}$.

\begin{figure}[ht]
\centerline{\epsfig{file=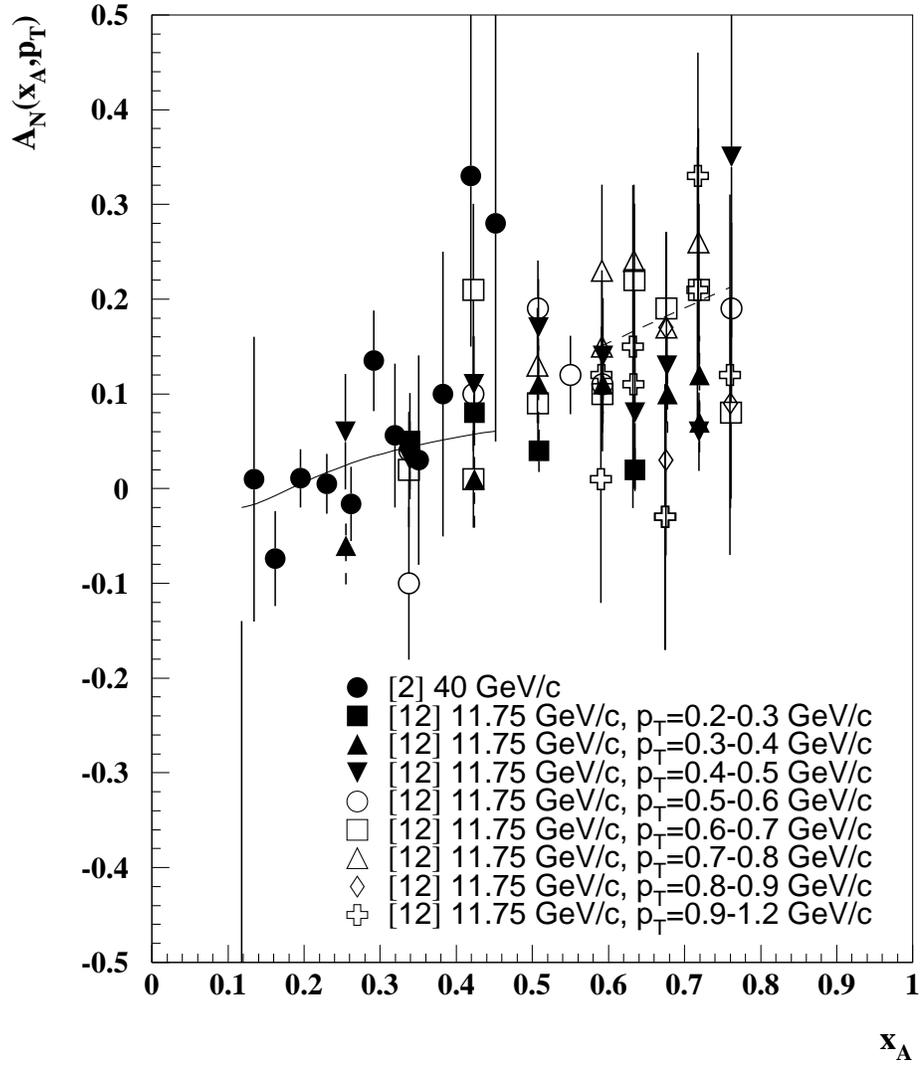,width=14cm}}
\caption {$\AN{}$ vs $\XA{}$ for the $K^{+}$ production  by polarized 
protons. The solid fitting curve corresponds to the 40 GeV/c data [2], and 
the dashed curve corresponds to the 11.75 GeV/c data [12] and 
$0.5 \le \PT{} \le 0.6$ GeV/c.}
    \label{k3pp_12xa}
\end{figure}

The analyzing power for $K^{+}$-meson production  in 
$p^{\uparrow} \!p$-collisions 
has been measured in  two experiments \cite{FODS,PRD18} at 40 and 11.75 GeV/c, 
respectively. It is shown in Fig.~\ref{k3pp_12xa}  as a function of $\XA{}$.
 The $\AN{}$ dependence  on kinematic variables was 
approximated by eq.~(\ref{eq:AN1}) with $a_{4}=0$ and $a_{6}=0$, because 
statistical accuracy of the data is limited. The fit parameters are presented 
in Table~\ref{KTAB1}. The experimental data are compatible with 
the $\XA{}$-scaling  (see eq.~(\ref{eq:AN1})).

\begin{figure}[ht]
\centerline{\epsfig{file=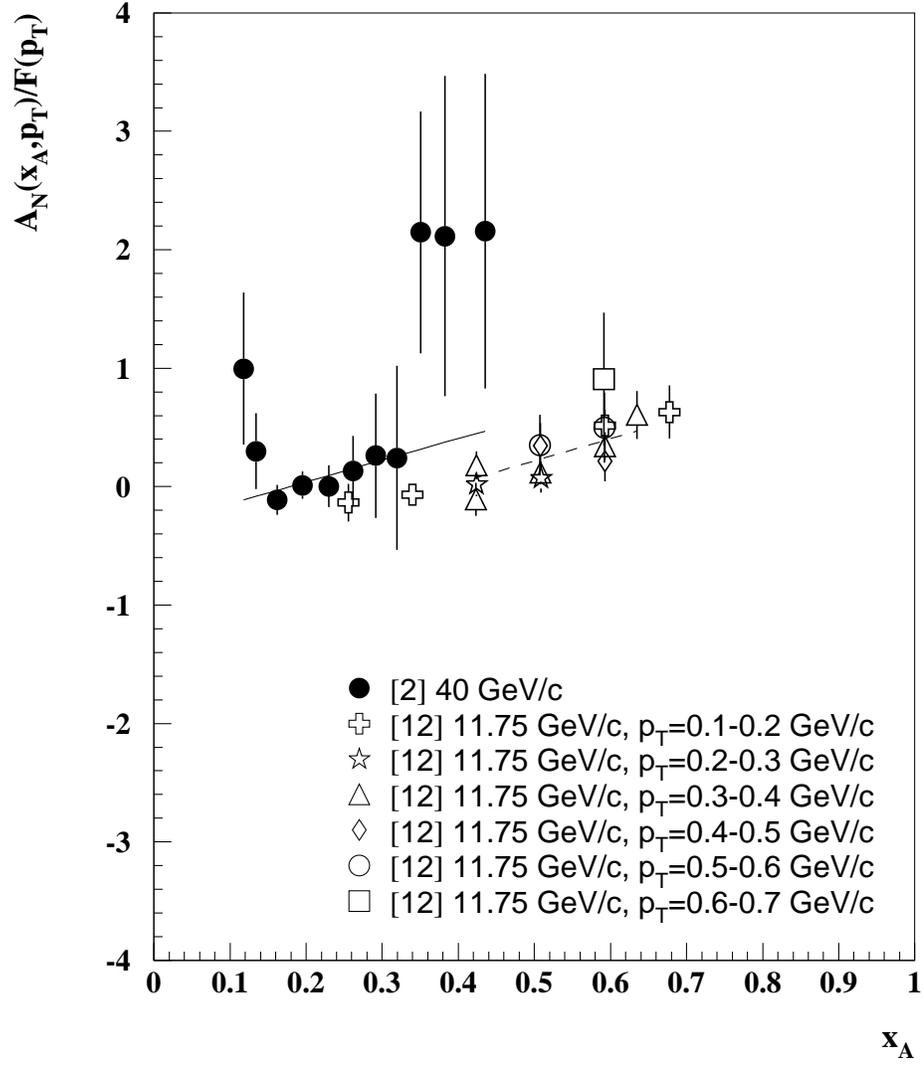,width=14cm}}
\caption {The ratio $\AN{}/{\FPT}(\PT{})$ vs $\XA{}$ for the $K^{-}$ %
production by polarized protons. The solid fitting curve corresponds to 
the data [2],  and the dashed curve corresponds  to the data [12] 
and region $0.3 \le \PT{} \le 0.4$ GeV/c.}
    \label{k4ra_13xa}
\end{figure}

 The analyzing power for $K^{-}$-meson production has been measured at 40
 and 11.75 GeV/c \cite{FODS,PRD18}. It was fitted by eq.~(\ref{eq:AN1}) with
$a_{6}$, as a free parameter and $a_{4}=0$. The energy dependent term $a_{6}/s$
significantly improves the fit for $K^{-}$, in contrast to the $K^{+}$ case.
The parameters of the fit are shown in Table~\ref{KTAB1}.
 The ratio  $\AN{}/{\FPT}(\PT{})$
is shown in Fig.~\ref{k4ra_13xa} vs $\XA{}$, 
where the shift of data points due to $a_{6}/s$ term is
clearly seen. The parameter $m$  for $K^{-}$-meson, which has no valence 
quarks common for colliding protons, is much smaller than in the case with
 $K^{+}$-meson and  is close to the estimation of Ref.~\cite{RYSK}. Contrary 
to  $\pi^{\pm}$-mesons, $K^{\pm}$-mesons do not show  any unusual behaviour
 at 11.75 GeV/c which requires an  additional contribution to the analyzing
 power similar to that given  by eq.~(\ref{eq:AN0}).

\begin{figure}[ht]
\centerline{\epsfig{file=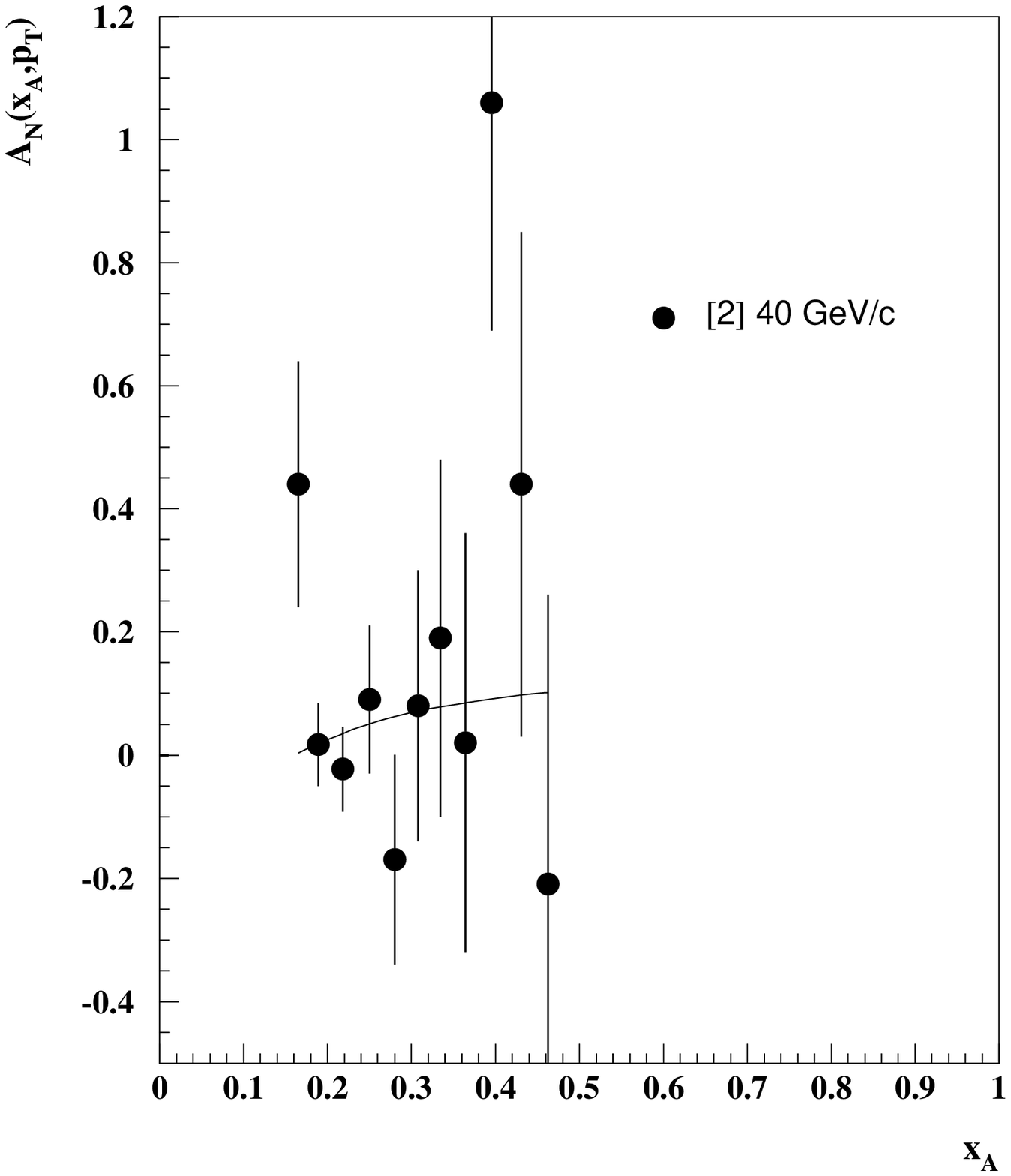,width=14cm}}
\caption {$\AN{}$ vs $\XA{}$ for antiproton %
 production by polarized protons. The curve corresponds to a fit 
by eq.~(6)  with the  parameters given in Table~5.}
     \label{p6pp_14xa}
\end{figure}

 The analyzing power for antiprotons has been measured only at 40 GeV/c at
 one fixed laboratory angle \cite{FODS}. Therefore, it is impossible
 to determine  parameter $m$, which was fixed at 1 GeV/c during the fit 
of the data by eq.~(\ref{eq:AN1}). The fit parameters are presented in
 Table~\ref{KTAB1} and $\AN{}$ vs $\XA{}$ is shown in Fig.~\ref{p6pp_14xa}.
  Additional measurements are required for $K^{+}$, $K^{-}$-mesons, and 
antiprotons at different energies and production angles to check 
the $\XA{}$-scaling and  determine the parameters of eq.~(\ref{eq:AN1}).
\begin{table}[htb]
\small
\vspace{1cm}
\caption{Fit parameters of eq.~(6) for the $\pi^{0}$,
 $K^+$, $K^-$-mesons and $\bar{p}$.
Parameters $a_{4}$ -- $a_{5}$ are set equal to zero and $\omega=1$
 during the fit, with $\epsilon= \pm 0.015$ for $\pi^{0}$ and 
$\epsilon= \pm 0.010$ for  $K^+$, $K^-$, $\bar{p}$. }
\begin{center}
\begin{tabular}{cccccc}    \hline
$h_{3}$ &  $c$        &      $x_{0}$     & $m$      & $a_{6}$
 & $\chi^{2}$ / points \\[0.3cm] \hline
$\pi^{0}$ & 0.24 $\pm$0.04  &  0.111$\pm$0.019 &1.40$\pm$0.49 & 0
& 50.5 / 54  \\[0.3cm] 
$K^+$ & 0.37 $\pm$0.08  &  0.183$\pm$0.045 &1.15$\pm$0.34 & 0
& 65.8 / 67  \\[0.3cm] 
$K^-$  & 1.88 $\pm$0.34  &  0.086$\pm$0.054 &0.25 $\pm$0.07 &-13.5 $\pm$4.2
& 24.2  / 28 \\[0.3cm] 
$\bar{p}$ & 0.6  $\pm$1.0   &  0.16 $\pm$0.12  &1.00          & 0
& 15.6 / 11 \\[0.3cm] \hline
 \end{tabular}
\end{center}
\label{KTAB1}
\end{table}
\section{Analyzing powers for $\Lambda$, $\KS$, $\eta$ production
 by polarized protons}
\begin{figure}[ht]
\centerline{\epsfig{file=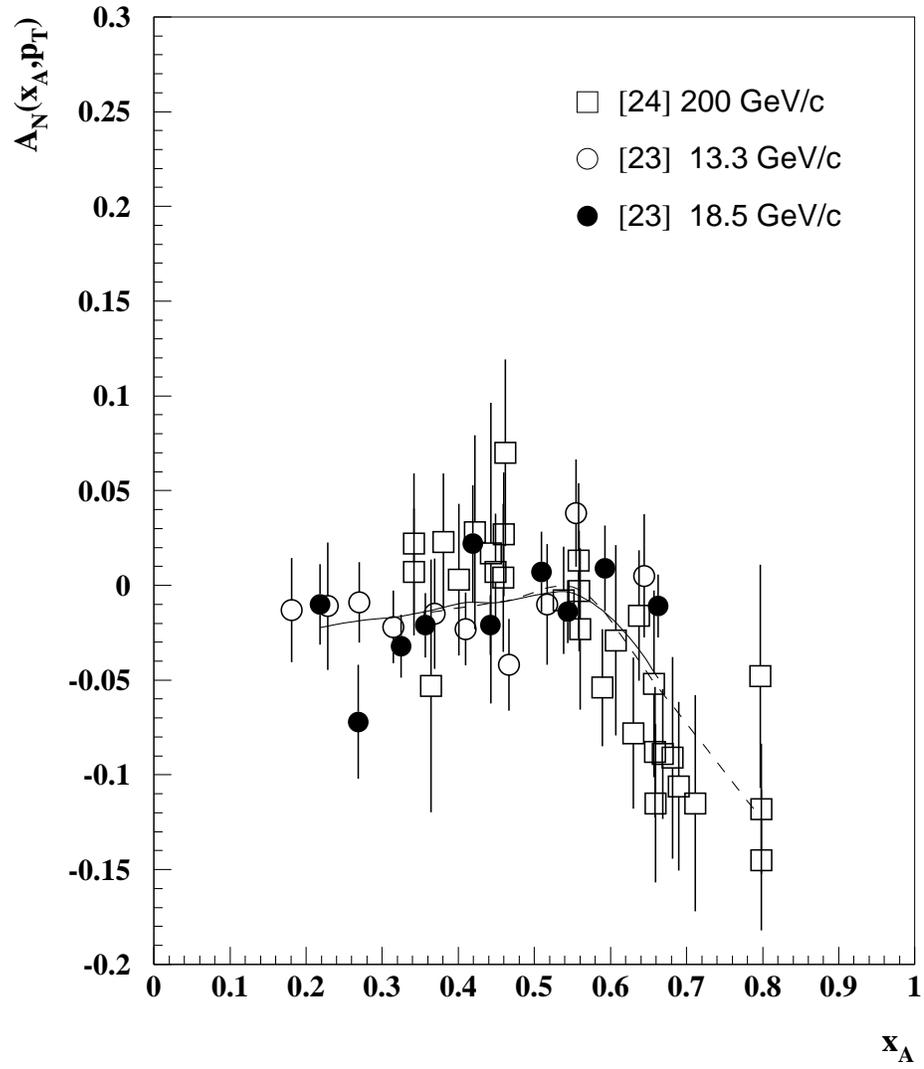,width=14cm}}
\caption {$\AN{}$ vs $\XA{}$ for the $\Lambda$ production %
by polarized protons.  The solid fitting curve corresponds to the 18.5 GeV/c 
data [23], and the dashed curve corresponds to the 200 GeV/c data [24].}
    \label{lampp_15xa}
\end{figure}

The analyzing power for the $\Lambda$-hyperon production has been measured 
at 13.3, 18.5 and 200 GeV/c \cite{PRD38,PRL75}. It is shown as a function 
of $\XA{}$ in Fig.~\ref{lampp_15xa} along with fitting curves 
(eq.~(\ref{eq:AN1})). Data  \cite{PRD38} were obtained on a Be target, 
and data  \cite{PRL75} on a proton target.
 The fit parameters for different $\XA{}$ definitions are presented in 
Table~\ref{TABlam}. The best  $\chi^{2}$ is attained  with $\XA{}$ defined
 by eq.~(\ref{eq:XFR}).
%
%
\begin{table}[htb]
\small
\caption{Fit parameters  of eq.~(6) for the $\Lambda$ and 
different definitions of scaling variable $\XA{}$, eqs.~(2) --(5),
with $\epsilon= \pm 0.015$ and $\omega = 1$.}
\begin{center}
\begin{tabular}{ccccccc}    \hline
 eq. &  $c$        &      $x_{0}$     & $m$      & $a_{4}$
& $a_{5}$       & $\chi^{2}$ / points \\[0.3cm] \hline
 (2)  &-0.52 $\pm$0.15  &  0.557$\pm$0.036 &0.66  $\pm$0.36  & 0.563$\pm$0.035
&-0.111$\pm$0.096&  39.4 / 49 \\[0.3cm] 
 (3) &-0.72  $\pm$0.38  &  0.539$\pm$0.021 &1.6  $\pm$1.3    & 0.527$\pm$0.024
&-0.158 $\pm$ 0.073&  24.3 / 49  \\[0.3cm] 
 (4)    &-0.54 $\pm$0.15  &  0.560$\pm$0.034 &0.69 $\pm$0.37 & 0.564$\pm$0.033
&-0.109$\pm$0.091&  38.3 / 49  \\[0.3cm] 
 (5)   &-0.53 $\pm$0.15  &  0.559$\pm$0.034 &0.68  $\pm$0.37 & 0.564$\pm$0.034
&-0.109$\pm$0.091&  38.5 / 49 \\[0.3cm] \hline
 \end{tabular}
\end{center}
\label{TABlam}
\end{table}     
  As is seen from Fig.~\ref{lampp_15xa}, $\AN{}$ can be described at different
energies by the same function of the scaling variable $\XA{}$ at the present
level of experimental errors. The analyzing power is close to zero for the
 region $0.2 \le \XA{} \le 0.6$ and is negative for the $\XA{}$ above 0.6. 

\begin{figure}[ht]
\centerline{\epsfig{file=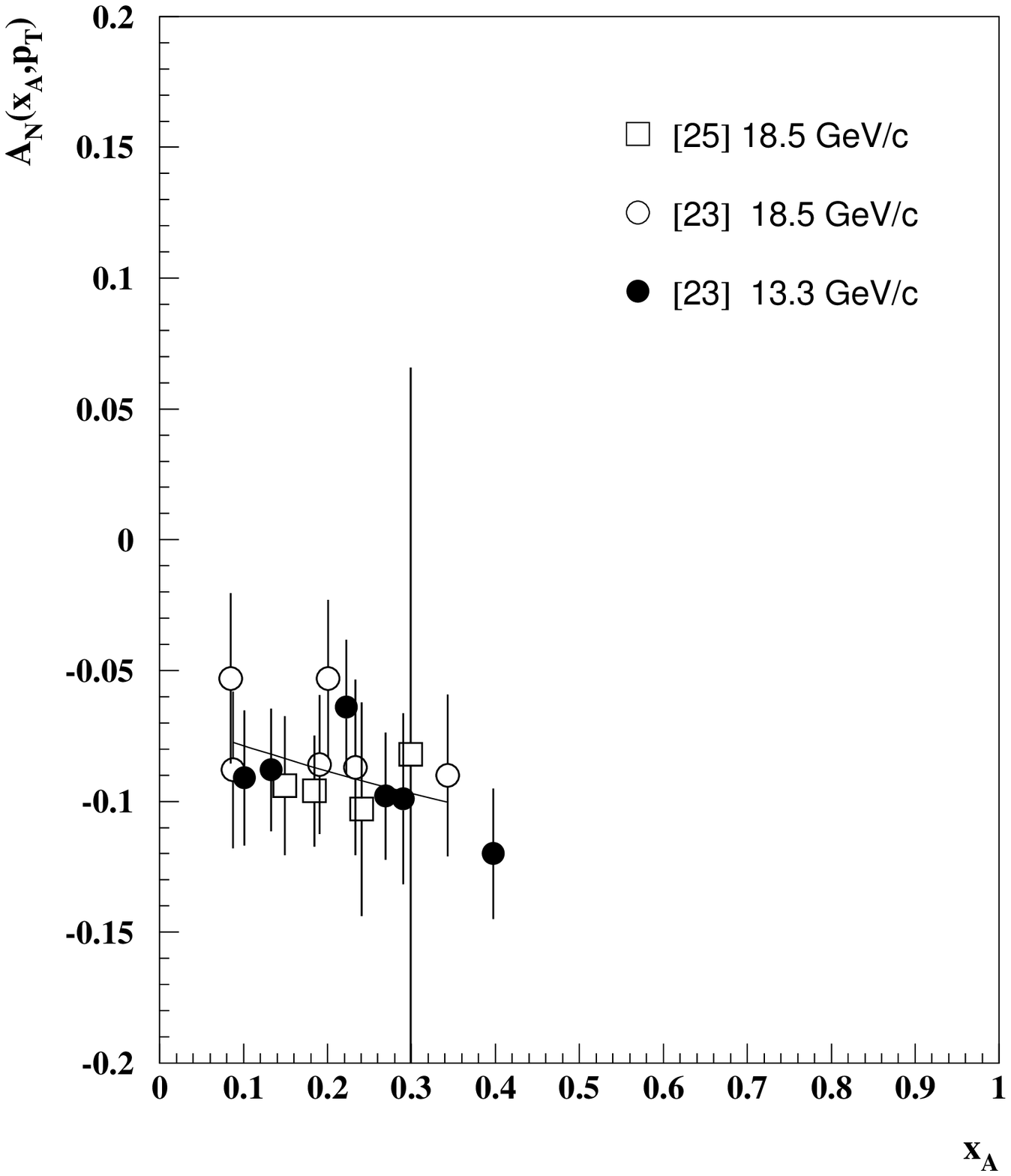,width=14cm}}
\caption {$\AN{}$ vs $\XA{}$ for the $\KS$ production %
by polarized protons.  The fitting curve corresponds to 
the 18.5 GeV/c data [23].}
    \label{kspp_16xa}
\end{figure}

  Measurements of $\AN{}$ for the $\KS$-mesons have been performed at 
13.3  and 18.5 GeV in the central region only \cite{PRD38,PRD41}, both on a Be
target. 
In Fig.~\ref{kspp_16xa} $\AN{}$ is shown as a
 function of $\XA{}$  along with a fitting 
curve given by eq.~(\ref{eq:AN1}). The fit parameters are presented 
in Table~\ref{KSTAB}. The data are compatible with the $\XA{}$-scaling, 
but additional measurements are desirable to check it at different energies 
and in the fragmentation region.
\begin{table}[htb]
\small
\caption{Fit parameters  of eq.~(6) for the $\KS$ and 
$\eta$-mesons, with $\epsilon= \pm 0.015$ and $\omega=1$. }
\begin{center}
\begin{tabular}{cccccc}    \hline
$h_{3}$     &  $c$        &      $x_{0}$     & $m$   
 & $\chi^{2}$ / points \\[0.3cm] \hline
$\KS$    & -0.143$\pm$0.095 & -0.49 $\pm$0.50  &0.79  $\pm$0.49  
 &   4.4 / 16 \\[0.3cm] 
$\eta$ &  1.00 $\pm$0.36  &  0.323$\pm$0.048 &1.00             
 & 0.0  / 4  \\[0.3cm] \hline
 \end{tabular}
\end{center}
\label{KSTAB}
\end{table}     

\begin{figure}[ht]
\centerline{\epsfig{file=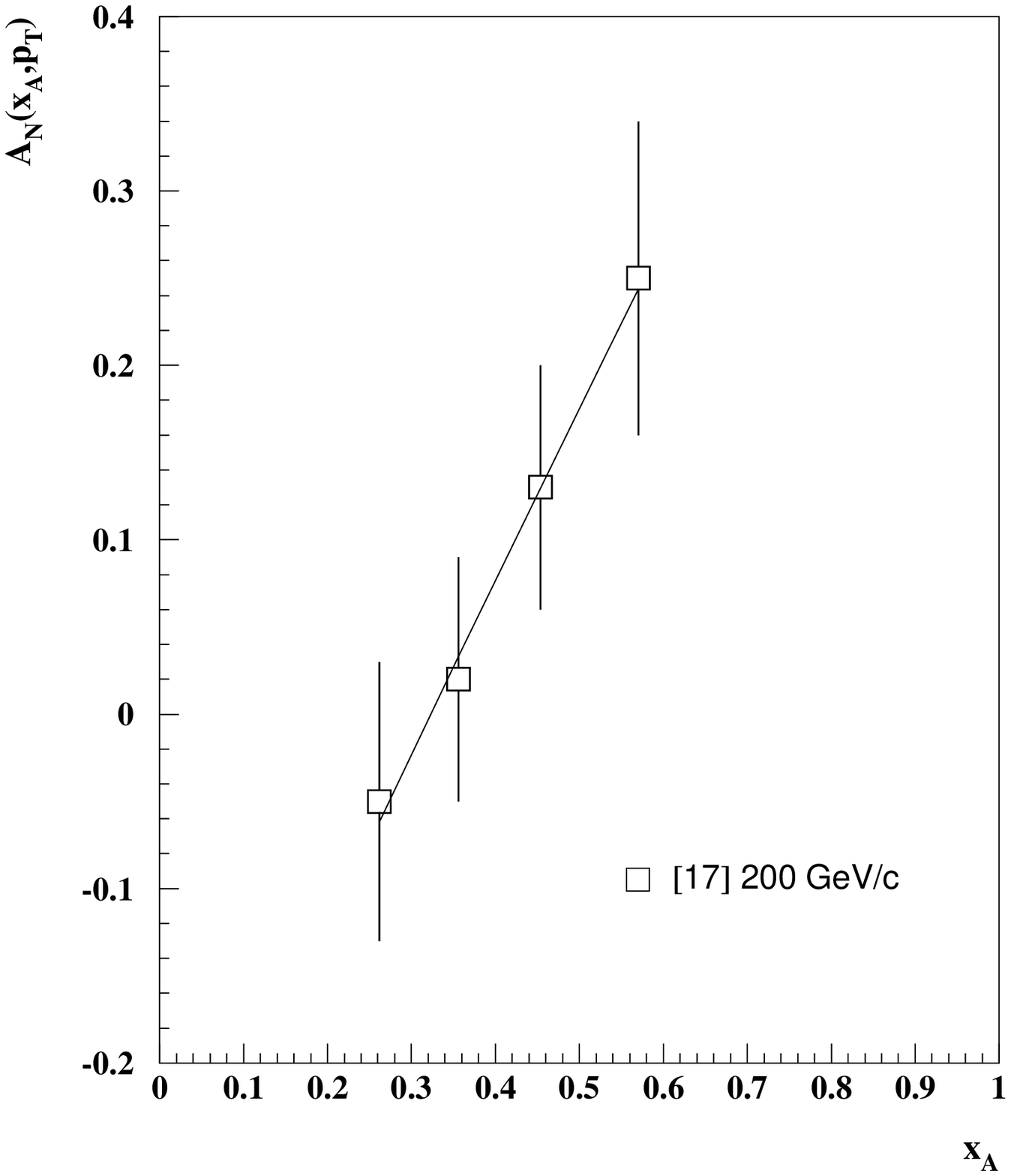,width=14cm}}
\caption {$\AN{}$ vs $\XA{}$ for the $\eta$ production %
by polarized protons. The curve corresponds to a fit (6) with the parameters
given in Table~7.}
    \label{etapp_17xa}
\end{figure}

 The analyzing power for the $\eta$-meson production in $p^{\uparrow} \!p$ 
collisions has  been measured at 200 GeV/c \cite{AD9756}. It is shown in 
Fig.~\ref{etapp_17xa}  along with  the fitting curve,  eq.~(\ref{eq:AN1}). 
The fit parameters are shown in  Table~\ref{KSTAB}. Since the measurement 
has been performed  at a fixed angle, parameter $m$ was  fixed during 
the fit.

\section{Analyzing powers for the $\pi^{\pm}$, $\pi^{0}$ and $\eta$ 
production in $\bar{p}^{\uparrow} \!p$ collisions}
\begin{figure}[ht]
\centerline{\epsfig{file=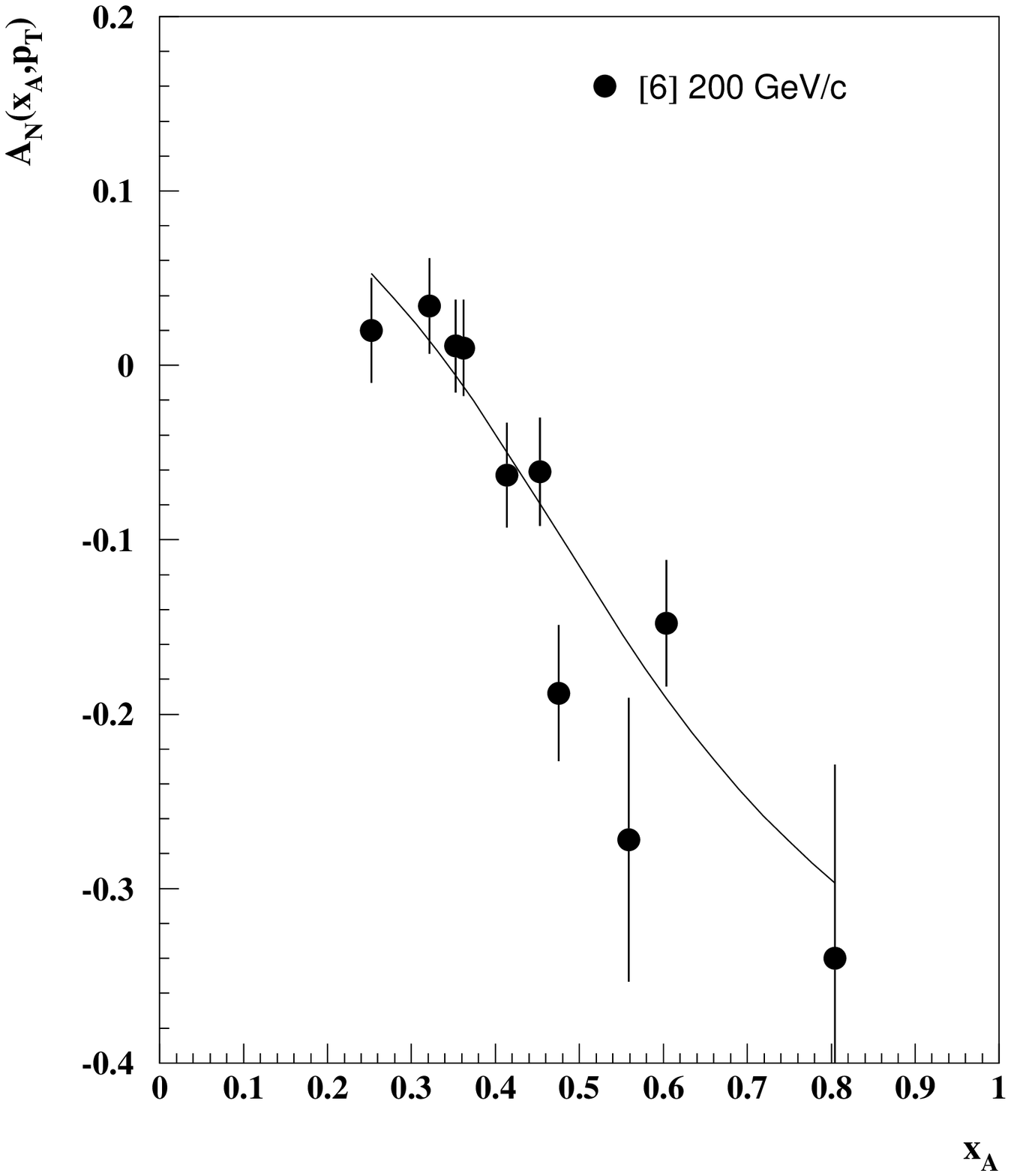,width=14cm}}
\caption {$\AN{}$ vs $\XA{}$ for the $\pi^{+}$ production %
in $\bar{p}^{\uparrow} \!p$-collisions. The curve corresponds to 
a fit by eq.~(6) with the parameters given in Table~9.}
     \label{pi1ap_20xa}
\end{figure}

\begin{figure}[ht]
\centerline{\epsfig{file=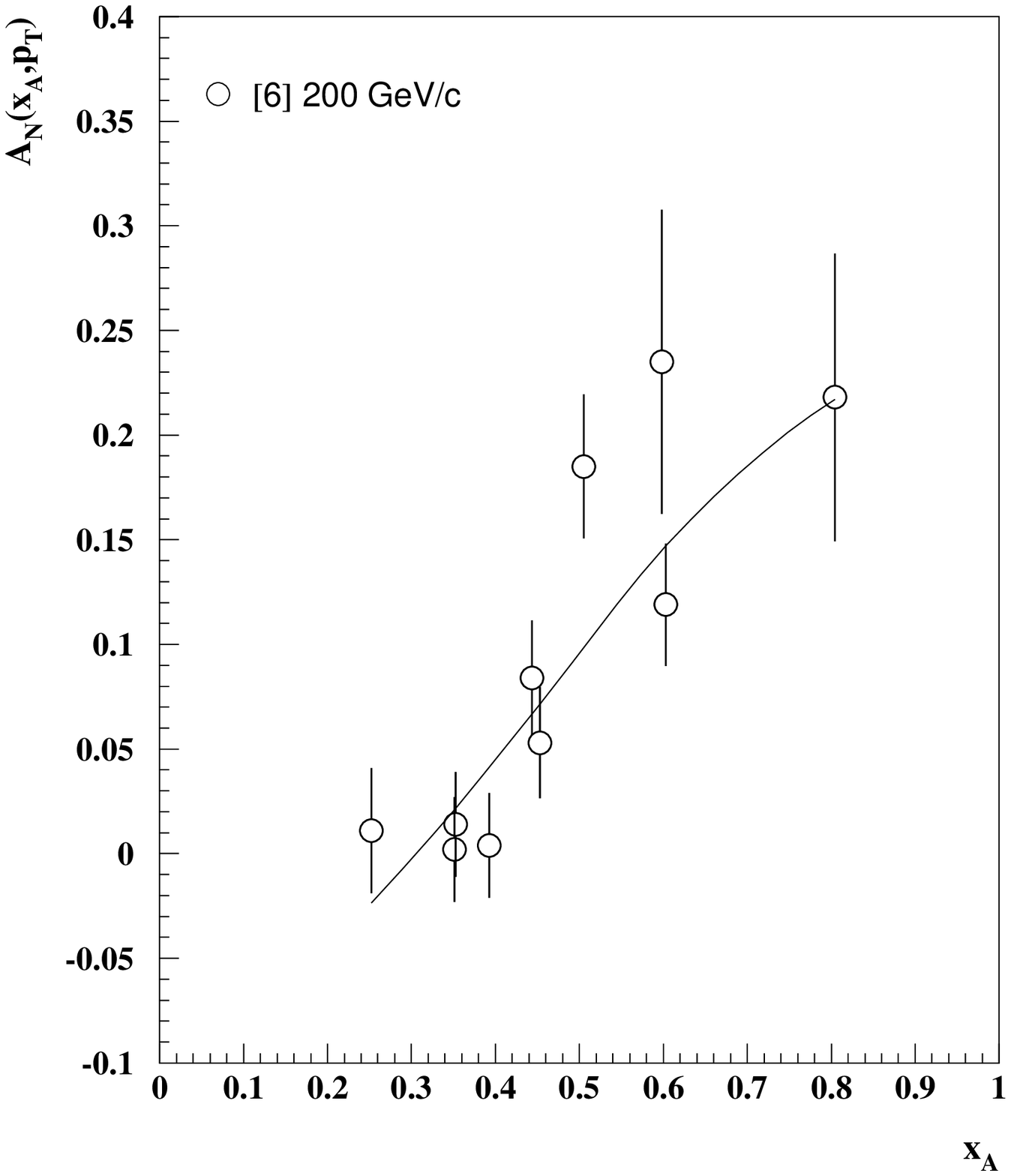,width=14cm}}
\caption {$\AN{}$ vs $\XA{}$ for the $\pi^{-}$ production %
in $\bar{p}^{\uparrow} \!p$-collisions. The curve corresponds to 
a fit by eq.~(6) with the parameters given in Table~9.}
     \label{pi2ap_21xa}
\end{figure}

  The analyzing power for the $\pi^{\pm}$-meson production in the 
fragmentation region of polarized antiprotons has been measured at
 200 GeV/c \cite{PRL77}.
 It is shown in Figs.~\ref{pi1ap_20xa}
 and \ref{pi2ap_21xa}, as a function of $\XA{}$, 
for the $\pi^{+}$ and $\pi^{-}$,
respectively. The fit parameters are presented in Table~\ref{pi_ap}. Parameter
$m$  has been fixed due to limited statistics.
%
\begin{table}[htb]
\small
\caption{Fit parameters  of eq.~(6) for the $\pi^{\pm}$, 
$\pi^{0}$, and $\eta$-meson production in $\bar{p}p$-collisions,
with $\epsilon= \pm 0.015$. }
\begin{center}
\begin{tabular}{cccccc}    \hline
$h_{3}$       &  $c$         &      $x_{0}$     & $m$  &  $\omega$
  & $\chi^{2}$ / points \\[0.3cm] \hline
$\pi^{+}$    & -0.32 $\pm$0.20  & 0.344$\pm$0.020 &1.0        & 2.8$\pm$2.1
  & 10.4 / 10 \\[0.3cm] 
$\pi^{-}$    &  0.23 $\pm$0.10  & 0.309$\pm$0.035 &1.0        & 2.8$\pm$1.8
  & 10.1 / 10 \\[0.3cm] 
$\pi^{0}$    &  0.15 $\pm$0.07  & 0.050$\pm$0.061 &1.5 $\pm$1.3 & 1.0     
  & 21.1 / 34 \\[0.3cm] 
$\eta$    & -1.1  $\pm$0.9   & 0.468$\pm$0.075 &1.0          & 1.0          
  &  0.9 / 3  \\[0.3cm] \hline
 \end{tabular}
\end{center}
\label{pi_ap}
\end{table}                                 

\begin{figure}[ht]
\centerline{\epsfig{file=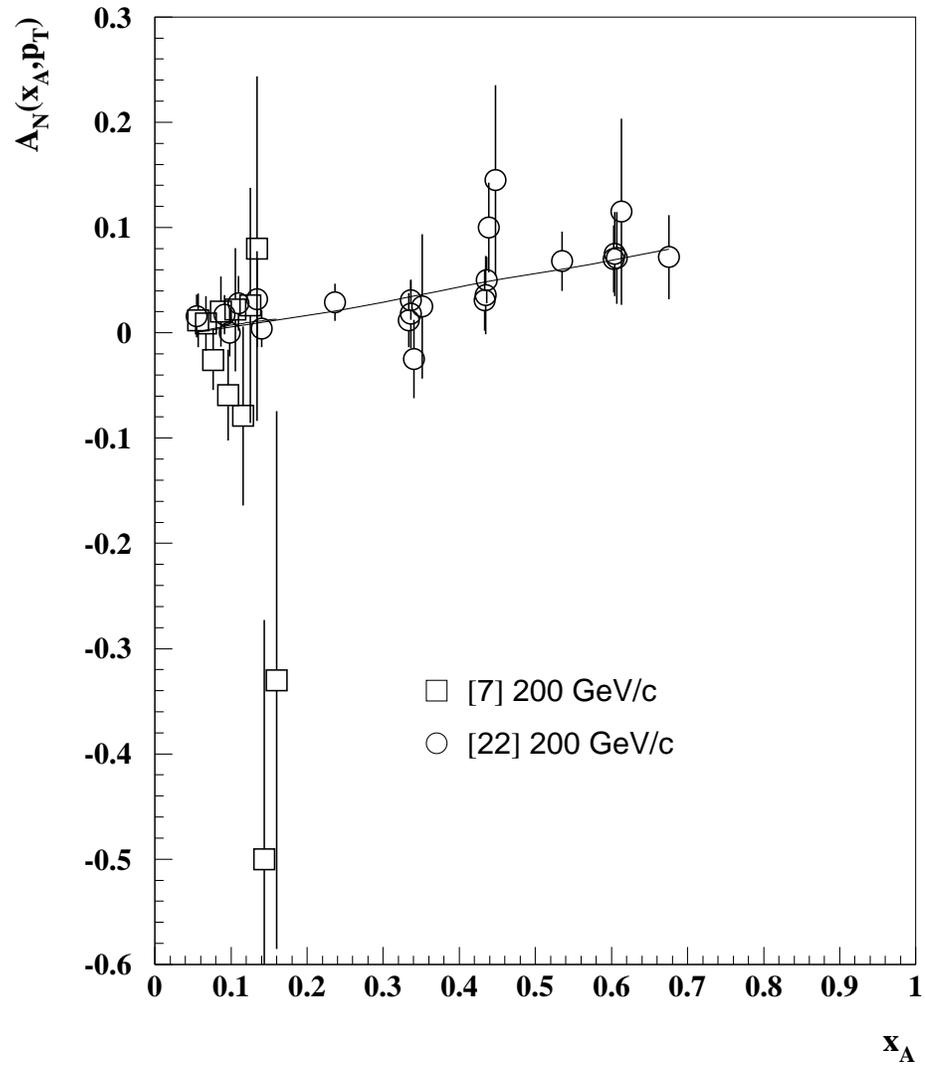,width=14cm}}
\caption {$\AN{}$ vs $\XA{}$ for the $\pi^{0}$ production %
in $\bar{p}^{\uparrow} \!p$-collisions. The fitting curve corresponds to
the 200 GeV/c data [22].}
     \label{pi7ap_22xa}
\end{figure}

 Measurements of $\AN{}$ for the $\pi^{0}$-meson production in 
$\bar{p}^{\uparrow} \!p$-collisions 
has been performed at 200 GeV/c in the central region \cite{PRD53} and the
fragmentation  region \cite{PL261} of polarized antiprotons. The data are 
shown  as a function of $\XA{}$ along with the fitting curve
 (eq.~(\ref{eq:AN1})) in Fig.~\ref{pi7ap_22xa}. 
%
The fit parameters are shown in Table~\ref{pi_ap}.  As in the case of
 polarized proton beam, high $\PT{}$ data do not show any significant 
analyzing power, in agreement with  the predictions of $\XA{}$-scaling.

 The analyzing power for the $\eta$-meson production has been measured just 
in a few points at 200 GeV/c \cite{AD9756}. The fit parameters 
are shown in Table~\ref{pi_ap}. 

It is easy to notice that $x_{0}$-parameter 
 for the $\pi^{\pm}$ and 
$\eta$-meson production by polarized antiprotons is by about 0.15 larger 
as compared to the case of polarized proton beam.
\section{Asymmetries for the $\pi^{0}$ and $\eta$ production in
 $\pi^{-}p^{\uparrow} $ collisions}
\begin{figure}[h]
\centerline{\epsfig{file=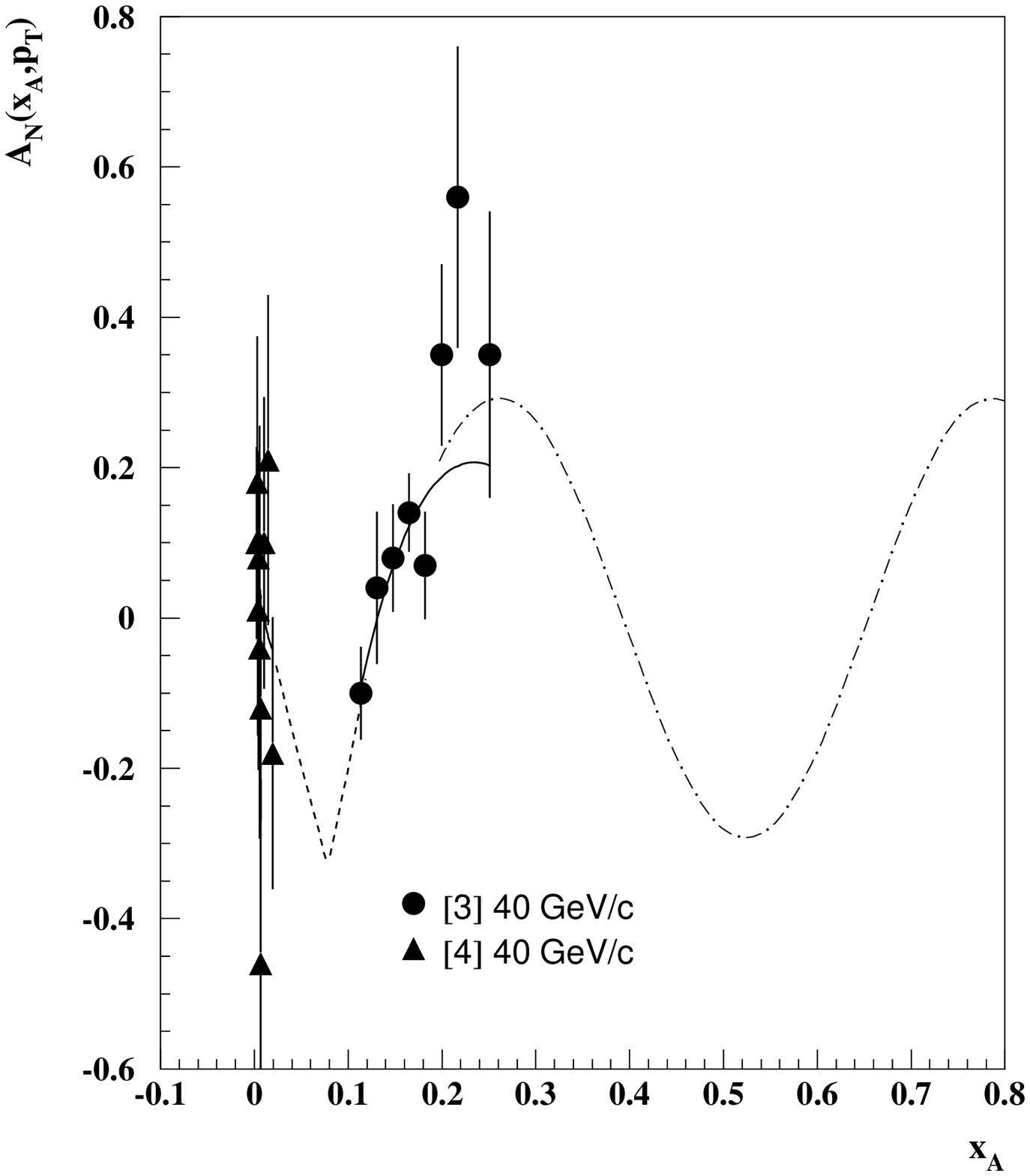,width=14cm}}
\caption { $\AN{}$ vs $\XA{}$ for the $\pi^{0}$ production 
in  $\pi^{-}p^{\uparrow}$ -collisions. The solid curve corresponds 
to a fit by eq.~(6) with the parameters given in Table~8.
The dashed curve corresponds to an extrapolation
of the fit (6) for the region $\PT{}$=1 GeV/c and $0.03 \le \XA{} \le 0.1$.
The dash-dot curve corresponds to an extrapolation
of the fit (6) for the region $\PT{}$=2 GeV/c and $ \XA{} \ge 0.3$.}
    \label{pi7pi_18xa}
\end{figure}

   Asymmetry measurements for the $\pi^{0}$ and $\eta$-meson production
have been carried out at 40 GeV/c in the central region \cite{Amag}
and in the fragmentation region of $\pi^{-}$-meson \cite{APOK49}.
The data for the $\pi^{0}$-mesons are shown in Fig.~\ref{pi7pi_18xa}
 along with the fitting curve (eq.~(\ref{eq:AN1})). The dashed curve shows 
the prediction  of eq.~(\ref{eq:AN1}) for region  
$0.03 \le \XA{} \le 0.1$ and  $\PT{}=1$ GeV/c, where no data exist and 
a local minimum of $\AN{}$ is expected from the fit. 
The dash-dot curve shows the prediction for region $\PT{}=2$ GeV/c and  
$\XA{} \ge 0.3$, where a local maximum of $\AN{}$ is expected near $\XA{}=0.3$,
and a local minimum is expected near $\XA{}=0.5$.
The use of  $\sin (\omega(\XA{}-x_{0}))$ with large $\omega$ value
in eq.~(\ref{eq:AN1}) allows one to satisfy  the constrain $|\AN{}| \le 1$. 
The fit parameters are shown in Table~\ref{pibeam}. The values of $\omega$ 
shown in Table~\ref{pibeam} are the minimal ones which satisfy  the constrain
$|\AN{}| \le 1$. 
\begin{table}[htb]
\small
\caption{Fit parameters  of eq.~(6) for the $\pi^{0}$ 
and $\eta$-meson production in $\pi^{-}p$~-collisions, with $\epsilon=0.015$.}
\begin{center}
\begin{tabular}{cccccccc}    \hline
 $h_{3}$       &  $c$         &      $x_{0}$     & $m$  &  $a_{4}$
& $a_{5}$    & $\omega$         & $\chi^{2}$ / points \\[0.3cm] \hline
$\pi^{0}$  &  1.0 $\pm$0.4 & 0.131$\pm$0.008 &0.3 $\pm$1.3 &0.078$\pm$0.062
&-0.8$\pm$1.7&  12.0           & 15.2 / 20 \\[0.3cm] 
$\eta$   &  1.0 $\pm$0.5  & 0.154$\pm$0.016 &1.0          & 0.0          
&    0.0     &12.5              &  0.2 / 3  \\[0.3cm] \hline
 \end{tabular}
\end{center}
\label{pibeam}
\end{table}                                 
  
\begin{figure}[h]
\centerline{\epsfig{file=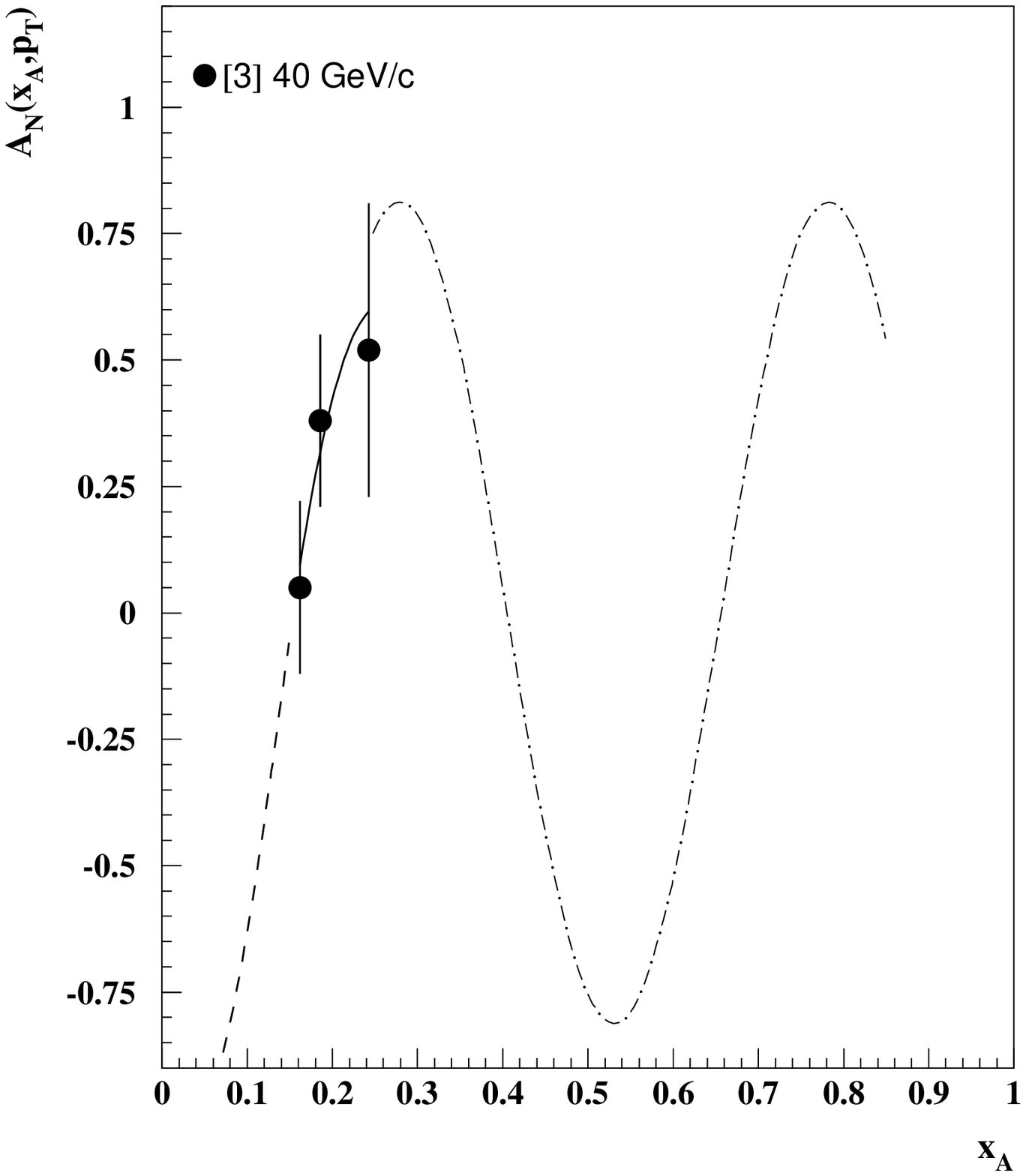,width=14cm}}
\caption { $\AN{}$ vs $\XA{}$ for the $\eta$ production %
in $\pi^{-}p^{\uparrow}$-collisions.  The solid curve corresponds to 
a fit by eq.~(6) with the parameters given in Table~8. 
The dashed curve corresponds to an extrapolation
of the fit (6) for the region $\PT{}$=1 GeV/c and $ \XA{} \le 0.15$.
The dash-dot curve corresponds to an extrapolation
of the fit (6) for the region $\PT{}$=2 GeV/c and $ \XA{} \ge 0.3$.}

    \label{etapi_19xa}
\end{figure}

%
  The asymmetry vs $\XA{}$ for the $\eta$-meson production is shown in 
Fig.~\ref{etapi_19xa}. Since only a few experimental points have been 
measured,  some of the parameters of  eq.~(\ref{eq:AN1}) were fixed
(see Table~\ref{pibeam}). 
Predictions for region $\XA{} \le 0.15$ and $\PT{}$ = 1 GeV/c are shown by
the dashed curve, and predictions for region 
 $\XA{} \ge 0.3$ and $\PT{}=2$ GeV/c are shown in Fig.~\ref{etapi_19xa}
 by the dash-dot curve. As in the case of $\pi^{0}$ production, a 
local maximum of $\AN{}$ is expected (near $\XA{}=0.3$). Also a local minimum
of $\AN{}$ is expected near $\XA{}=0.5$. 
For both $\pi^{0}$ and $\eta$-meson production by $\pi^{-}$
beam, the dependence on $\XA{}$  looks very similar having a fast rise in 
the range  $0.15 \le \XA{} \le 0.3$. This behaviour is very different from the
 $\XA{}$-dependence in $p^{\uparrow} \!p$-collisions,
where the rise of $\AN{}$ with $\XA{}$ is not so dramatic and $sin(x)$ function
in eq.~(\ref{eq:AN1}) is not very important at the present level of accuracy.
\section{Discussion}
  In this section we will try to understand the observed  $\XA{}$-scaling, 
which is approximated by eqs.~(\ref{eq:AN1}) -- (\ref{eq:AN}), within
 the framework of the ideas of existing models. We begin our discussion of the
results with a set of rules which reproduce the known features of the data.

The analyzing power for hadron production, as well as hyperon 
polarization in inclusive reactions are proportional to an imaginary part of 
the product of spin-flip  and spin-nonflip amplitudes
\begin{equation}
  \AN{} \propto \im (f_{\SNF}f_{\SF}^{*}) = 
  |f_{\SNF}||f_{\SF}|sin(\Delta \phi), \quad   \label{eq:ImFsf}
\end{equation}
where $\Delta \phi$ is a phase difference of the corresponding amplitudes
 \cite{Amag,RYSK,CRAI}. The equality of $\Delta \phi$ to zero means  $\AN{}=0$,
so we may suggest that at $\XA{}=x_{0}$ phase difference $\Delta \phi=0$
in case of $\pi^{+}$-meson production at high energy and $\PT{}$.

 The sign of analyzing power at a quark level is given by the rule: A quark
 with spin $upward$ prefers scattering to the $left$, and vice versa.
 Such result is easy to  get  by taking into account the interaction of a
 quark chromomagnetic momentum  with  chromomagnetic field, arising after 
the collision during hadronization \cite{RYSK}. This rule is also a direct
 consequence of the experimental observations \cite{LIANG}.

  The effect of recombination of  partons in the proton while they 
transfer into an outgoing hadron may be different depending on whether
they are accelerated (as with slow sea quarks) or decelerated (as with
 fast valence quarks).
  Slow partons mostly recombine with their spin downwards in the scattering
plane while fast partons recombine with their spin upward \cite{PRD24}.

  The existence of the  $x_{0}$ point in eq.~(\ref{eq:AN1}),
 where the analyzing power changes its sign,
can be explained by the same arguments which are used to explain the 
$\xf{}$-dependence of $\Lambda$-hyperon polarization in the SU(6) based parton
recombination model \cite{PRD24}. Following the same arguments we can say that
 the analyzing power for  $\Lambda$-production is proportional to
 $\Delta p$-change  in the momentum of sea $s$-quark:
\begin{equation}
  {\Delta \ps} \propto 1/3(\xf{} - 3\xs), \quad  \label{eq:DPlam}
\end{equation}
where $\xs \approx 0.1$ is a fraction of proton momentum, 
which carries sea $s$-quark. We assume here that the above rules concerning
close relation of quark polarization and analyzing power of scattering are
 valid. Substituting $\xf{}$ by $\XA{}$,  we get the expression similar
 to eq.~(\ref{eq:AN1}) with $x_{0}=3\xs$ about 0.3, which agrees 
qualitatively with the experimental  data  (see Fig.~\ref{lampp_15xa}) 
for the production analyzing power of $\Lambda$-hyperon, which is close to
 zero for $0.2\le \XA{} \le 0.6$. The only  difference consists in the
 absence of $sin(x)$  function in eq.~(\ref{eq:DPlam}), which is not very
 essential since the analyzing power is small.

  In case of $\pi^{+}$, $K^{+}$-meson production we can apply similar 
arguments. In this case $\Delta p$ for sea  quark ($\bar{d}$ or $\bar{s}$) is
equal to
\begin{equation}
  \Delta p_{SEA} \propto 1/2(\xf{} - 2x_{SEA}), 
\quad  \label{eq:DPpi}
\end{equation}
and we again have the expression similar to eq.~(\ref{eq:AN1}) with 
$x_{0}=2x_{SEA} $ about 0.2 in agreement with the experimental data 
(see Table~\ref{Tab1pi+}). An accelerated sea quark has spin downwards
and recombines with a valence spin upward $u$-quark from a polarized proton, 
producing $\pi^{+}$ or $K^{+}$-meson preferably to the left, which means 
a positive analyzing power. At $\XA{} \le x_{0}$, the acceleration is replaced
 by the deceleration, which reverses the sea and valence quark spin directions
 and the analyzing power sign.

 A dynamical reason for the above mentioned spin-momentum correlation is 
explained  in \cite{PRD24} by
the effect of Thomas precession \cite{Thom,Logunov}.  Another explanation of
spin-momentum correlation follows from a picture of a colour flux tube, which
emerges after the collision between an outgoing quark and the rest of hadronic
system \cite{RYSK,Ander}.

The analyzing power of $\pi^{+}$ production by polarized protons is 
determined by a  product of the elementary subprocess analyzing power
 ($\Aq$ for polarized quark  production), the polarization of this quark 
($\Pq$), and a ``dilution'' factor due to the presence of other contributions,
 not related with the valence quark fragmentation \cite{RYSK}
\begin{equation}
  \AN{} = \Aq\Pq\sigma(q)/( \sigma(q) + \sigma(g) ). \label{eq:ANRYS}
\end{equation}
The $u$-quark polarization according to SLAC~\cite{SLAC}, CERN~\cite{SMC} and
DESY~\cite{HERMES} measurements is positive and grows
with a fraction of momentum carried by quark and in the first approximation can
 be taken as $\Pq = \XA{}$, which is a generalization of $\Pq = \xf{}$,
assumed in \cite{RYSK}. For $\Aq$ we take the expression 
\begin{equation}
  \Aq = \delta \PT{} \cdot 2\PT{}/(m^{2} + \PT{}^{2}), \label{eq:AQ}
\end{equation}
where $\delta \PT{}$ ($\sim$ 0.1 GeV/c) is an additional transverse momentum, 
which quark with spin upward acquires in the chromomagnetic field of the flux
tube, and $m^{2}$ is some effective quark mass squared \cite{RYSK}.
 This expression for $\Aq$ is similar, in its functional form, to the lower 
order QCD calculations  and gives $\AN{}$ decreasing down to very small values
at very high $\PT{}$  \cite{RYSK,KANE}. In our case (eq.~(\ref{eq:AN1})) 
 $\Aq$ is proportional
 to ${\FPT}(\PT{})$,
given by eq.~(\ref{eq:F1}). The resulting expression for the $\AN{}$ is
\begin{equation}
  \AN{} = \delta \PT{}\cdot \XA{}\cdot 2\PT{}/(m^{2} + 
 \PT{}^{2}){D}(\XA{}), \quad     \label{eq:ANres}
\end{equation}
where  ${D}(\XA{})$ is a ``dilution'' factor mentioned above.
 Eq.~(\ref{eq:ANres}) is
very similar to eq.~(\ref{eq:AN1}) and to its high energy limit 
(\ref{eq:ANLIM})
 with $x_{0}=0$. The distinction consists in numerical values of parameters 
in eqs.~(\ref{eq:ANres}) 
and (\ref{eq:AN1}). In our case $\delta \PT{}=c\cdot m\cdot \omega=1.4$ GeV/c,
and $m = 2$ GeV, instead of $m=0.33$ GeV in \cite{RYSK}. We assume
here that the ``dilution'' factor ${D}(\XA{})$ is close to unity
at high $\XA{}$ values. The values of
the parameter $m$, obtained in  \cite{Amag} ($m=2$ GeV) turned out to be 
 much closer to that given in Table~\ref{Tab1pi+}.

   Another argument in favor of analyzing power and phase difference between 
spin-flip and spin-nonflip amplitudes to be proportional to hadron energy is 
given in \cite{Amag,AREST}. The reason is that the probability of quark 
spin-flip in an external field is proportional to a quark mean range before 
its hadronization. The experimental estimate of the hadronization range 
indicates that it is proportional to the secondary hadron energy \cite{FORML}.

 We may conclude that eq.~(\ref{eq:AN1}), which describes the scaling 
behaviour of analyzing powers, has a reasonable explanation of its basic
 components within the frameworks of existing models.

 Summarizing the above discussion we may assume that the observed 
$\XA{}$-scaling takes place due to the dependence of phase difference of 
spin-flip and spin-nonflip amplitudes at high $\PT{}$ and energy  on $\XA{}$ 
only.  This dependence for production of some hadrons 
($\pi^{+}, \pi^{0}, K^{\pm}, \KS, \eta, \bar{p}$) 
has a very simple form:
\begin{equation}
  \Delta \phi \propto \omega(\XA{}-x_{0}). \quad    \label{eq:DELphi}
\end{equation}

  The $x_{F}$-dependence (and hence the $\XA{}$-dependence)
 of the analyzing powers reflects in some models the corresponding dependence
 of the constituent quark polarization in the polarized proton \cite{TROSH2}.

 The $\PT{}$-dependence of the analyzing power, given by eq.~(\ref{eq:F1}),
 reflects  probably the ratio of spin-flip and spin-nonflip amplitudes 
\cite{Amag}:
\begin{equation}
  {\FPT}(\PT{}) = 2\PT{}m/(m^{2}+\PT{}^{2}) \propto
 { {|f_{\SNF}||f_{\SF}|}\over{|f_{\SNF}|^{2}+|f_{\SF}|^{2}}}. \label{eq:F1est}
\end{equation}  
 Both assumptions are not strictly proved, but they seem reasonable in view 
of the above stated arguments.

 It is interesting to note that maximum of ${\FPT}(\PT{})$ takes place  at 
about the same $\PT{}$, where the dip in elastic $p^{\uparrow} \!p$-scattering
  exists and where the interference maximum of spin-flip 
and spin-nonflip amplitudes takes place \cite{Fide}.

A more detailed comparison of different model predictions with the scaling 
behaviour of the experimental analyzing power is the subject for a separate
 paper.
\section{Possible application of inclusive reactions for the purpose
 of the beam polarimetry}
  A new generation of experiments with polarized proton beams requires
 a precise measurement of beam polarization. Unfortunately, above 100 GeV,
 the hadronic spin asymmetries used in most polarimeters are small and not 
well known.
   
  The Coulomb-nuclear interference (CNI) method has a systematic uncertainty
of the order of 10\% due to contribution of unknown hadronic spin-flip
amplitude~\cite{CNI}. The only experimental measurement of $A_{N}$ in
the CNI region ($-t \le 0.05$ GeV$^{2}$) at 200 GeV has relative errors
about 30\% or more~\cite{CNIEX}.

  The analyzing power of the Coulomb coherent process (the Primakoff effect)
 has been measured at 185 GeV polarized beam~\cite{CAREY}.
Relative experimental errors for the analyzing power were 21\% (statistical)
and 34\% (scale error due to the dilution factor), respectively. 
  
 Scaling properties of the analyzing power for the inclusive hadron production
 and its high value for some
of reactions allow, in principle, to use them for the purpose of
the beam polarimetry in a wide energy range. The most promising is
 the reaction of $\pi^{+}$  production in $p^{\uparrow}p$ or $p^{\uparrow}A$
 collisions, where $A$ is a light nucleus. Kinematic region 
 $p_{T} \ge 1$ GeV/c and $x_{A} \ge 0.5$  must be
used to achieve a reasonable relative accuracy (15\% or better).  This
accuracy is comparable with accuracy achieved using the analyzing power
of elastic $p^{\uparrow}p$ scattering, see for example \cite{E925}.
The agreement of the data~\cite{E925} on the carbon target with other data
on the proton target in Figs.~\ref{pi1pp_3xa}, \ref{pi1ra_6xa} and 
\ref{pi2pp_7xa} for $p_{T} \ge 0.8$ GeV/c supports a possible use of 
light nuclei targets in polarimeters.

  Other reactions with significant asymmetry in the region $x_{A} \ge 0.5$
and $p_{T} \ge 1$ GeV/c include $\pi^{-}$ and $\pi^{0}$ production in
 $p^{\uparrow}p$ or $p^{\uparrow}A$ collision. If a polarimeter is able
to identify different hadrons then all of them can be used to measure beam 
polarization and to decrease errors, both statistical and systematic.
  
 Further  improvement of the analyzing power experimental accuracy will
 make such polarimeters  competitive with other possibilities
 (e.g. the Primakoff effect, the elastic $p^{\uparrow}p$ scattering, etc).
\section{Scaling predictions for future experiments}
  The existence of the $\XA{}$-scaling is established from a limited set of
data which  cover only a restricted range of kinematic variables ($\PT{}$,
$\XA{}$, and $\sqrt{s}$). The corresponding c.m. production angles are 
concentrated mostly
near $0^{o}$, $90^{o}$, and $180^{o}$. A more detailed study of $\PT{}$,
$\XA{}$, and energy dependences could clarify theoretical basis for
the $\XA{}$-scaling and help to compare it with various models.

\begin{figure}[h]
\centerline{\epsfig{file=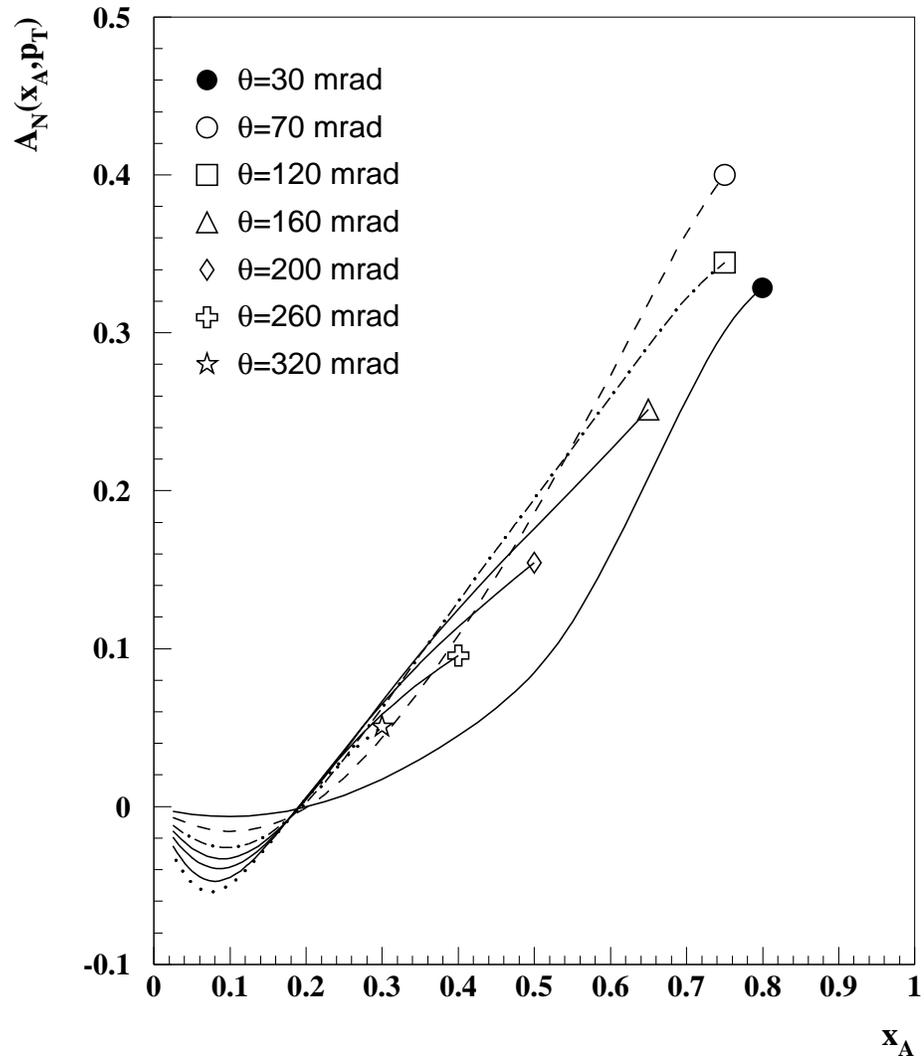,width=14cm}}
\caption {Predictions of the $\AN{}$ vs $\XA{}$ 
for the $\pi^{+}$ production by polarized 40 GeV/c protons %
at the different laboratory angles.}
     \label{pi1pp_23xa}
\end{figure}

  Detailed predictions of $\AN{}$ dependence on $\XA{}$  for 
$\pi^{+}$-production  at various
laboratory angles in $p^{\uparrow} \!p$ collisions at 40 GeV/c are shown
in Fig.~\ref{pi1pp_23xa}. This dependence is given by eqs.~(\ref{eq:AN1})
--(\ref{eq:AN}) with the parameters presented in Table~\ref{Tab1pi+}.
%
The measurements can be carried out at the FODS-2 experimental setup in IHEP
(Protvino) which uses a 40 GeV/c polarized proton beam \cite{FODS}. As is 
seen from Fig.~\ref{pi1pp_23xa}, the
 asymmetry is negative for the $\XA{}$ near 0.08 and grows in absolute
value with the increase of laboratory angle. At $\XA{}$ = 0.19 $\AN{}$ is 
always equal to zero and can be used to check systematic errors in the asymmetry
measurements. The largest values of $\AN{}$ are reached for laboratory
angle near 70 mrad. At this angle the values of $\XA{}$ and $\AN{}$ could be 
larger than 0.8 and 0.4, respectively. At smaller angles and large $\XA{}$, 
the asymmetry is  smaller owing to the decrease of $\PT{}$ and the 
corresponding reduction of the function ${\FPT}(\PT{})$ (see eq.~(\ref{eq:F1})).

\begin{figure}[h]
\centerline{\epsfig{file=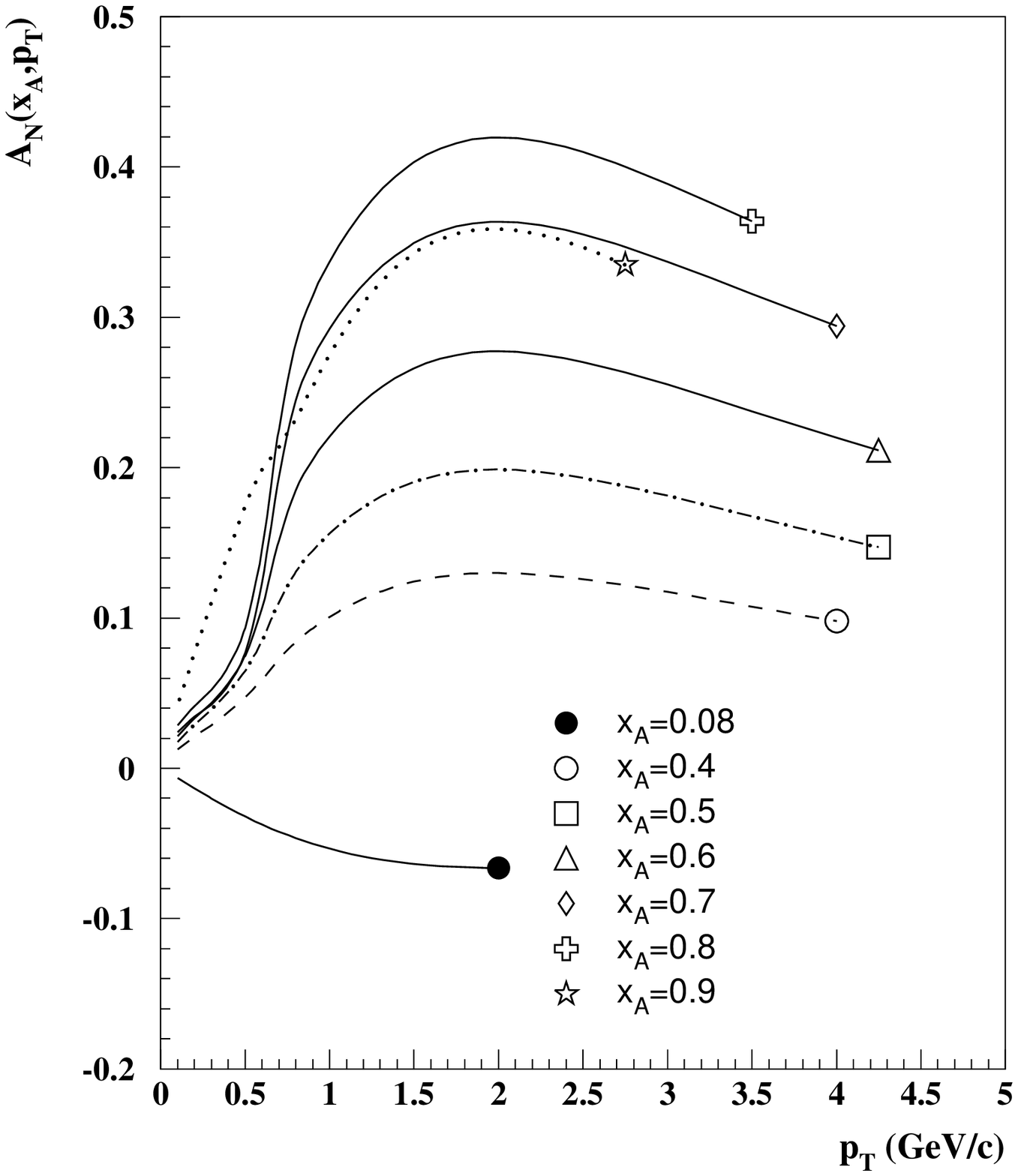,width=14cm}}
\caption {Predictions of the $\AN{}$ vs $\PT{}$ 
for the $\pi^{+}$ production by polarized 40 GeV/c protons %
at the different $\XA{}$ values.}
    \label{pi1pp_24xa}
\end{figure}

  The dependence of $\AN{}$  for the $\pi^{+}$ production in $p^{\uparrow} \!p$
collisions at 40 GeV/c on $\PT{}$ at several values of $\XA{}$ is shown in
Fig.~\ref{pi1pp_24xa}. It is possible to measure not only the rise of $\AN{}$
for  $0 \le \PT{} \le 2$ GeV/c, but also its probable decrease at higher
$\PT{}$ even at 40 GeV beam energy. For the $\XA{}$ values near 0.6, we can 
measure the shape of $\PT{}$-dependence up to 4 GeV/c.
 Much higher $\PT{} \le 10$ GeV/c
 can be reached at 200 GeV, where $\AN{}$ could decrease significantly 
(if there is no plateau at high  $\PT{}$) in
 comparision with its maximum value at $\PT{}$ about 2 GeV/c.

\begin{figure}[h]
\centerline{\epsfig{file=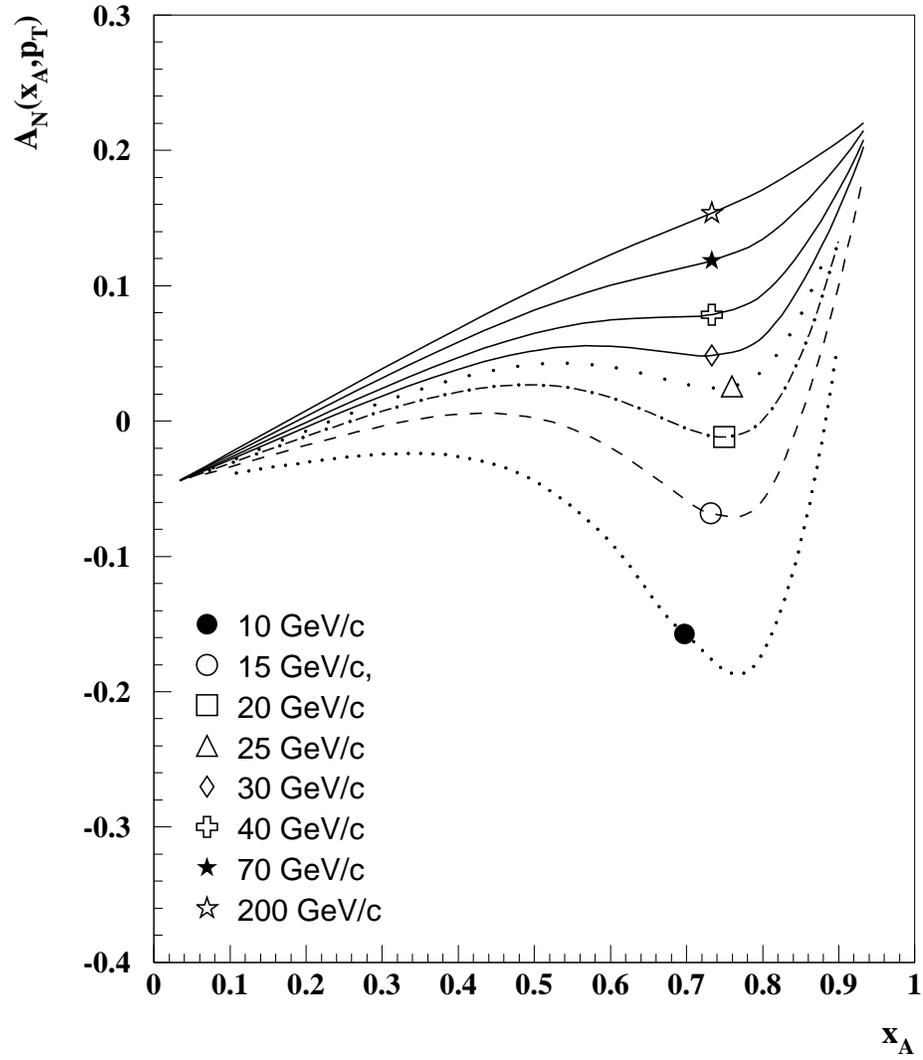,width=14cm}}
\caption {Predictions of the $\AN{}$ vs $\XA{}$ %
for the $\pi^{+}$ production by polarized protons %
at the different beam energies and $\PT{}=$ 0.5 GeV/c.}
    \label{pi1pp_25xa}
\end{figure}

The  $\AN{}$ dependence  for the $\pi^{+}$ production in $p^{\uparrow} \!p$ 
collisions  on $\XA{}$ at different beam energies and $\PT{}=$ 0.5 GeV/c
is shown in Fig.~\ref{pi1pp_25xa}. The corresponding parameters are taken
from Table~\ref{Tab1pi+}. As is seen in Fig.~\ref{pi1pp_25xa}, the asymmetry
approaches its high energy limit for the $\eb{} \ge$ 70 GeV. At smaller 
energies in the range from 10 to 40 GeV and the low $\PT{}$ value, 
there is a significant contribution
of nonasymptotic term (eq.~(\ref{eq:AN0}) ), which is most prominent at
$\XA{} = 0.75$, and changes the form of $\XA{}$-dependence. For energies
above 70 GeV the contribution of eq.~(\ref{eq:AN0}) is practically negligible.

  The oscillation of $\AN{}$ as a function of $\XA{}$ is predicted for the 
$\pi^{0}$  and $\eta$-meson production in $\pi^{-}p^{\uparrow}$-collisions 
(see Figs.~\ref{pi7pi_18xa} and \ref{etapi_19xa}).

  It is worth noticing that the expected maximum of $|\AN{}|$ for the 
$K^{+}$-meson production in $p^{\uparrow} \!p$-collisions is smaller than 
it is for the $\pi^{+}$-meson production. It could be related to a higher 
mass of  constituent $\bar{s}$-quark as compared with $\bar{d}$-quark mass. 
The higher constituent  quark mass leads to a smaller chromomagnetic momentum 
($\mu \propto 1/m_{q}$)  and  a smaller asymmetry \cite{RYSK}.

  A high value of $\AN{}$ is expected for the 
$p^{\uparrow} \!p \rightarrow K^{-} + X$ reaction at large $\XA{}$ 
 (see Table~\ref{KTAB1}) which 
contradicts some models that predict zero asymmetry \cite{ANSE}.

  We can make predictions for other reactions, using 
eqs.~(\ref{eq:AN1})--(\ref{eq:F1p}) and
the parameters, presented in Tables~\ref{Tab1pi+} -- \ref{pi_ap}.
\section{Conclusions}
It is shown that the existing analyzing power data   in
inclusive reactions for meson ($\pi^{\pm}$, $K^{\pm}$, $\KS$, $\eta$)
and baryon ($p$, $\bar{p}$, $\Lambda$) productions  
in $p^{\uparrow}\!p(A)$- and $\bar{p}^{\uparrow}\!p(A)$-collisions can
 be described by a simple function of
three variables ($\sqrt{s}$, $\PT{}$, $\XA{}$), where $\XA{}$ =  $E/\eb{}$
is a new scaling variable. In the limit of high enough energy ($\eb{} \ge$
40 GeV) and high $\PT{}$ ($\PT{} \ge$ 1.0 GeV/c), $\AN{}$ is a function of
$\XA{}$ and $\PT{}$ only with a precision of about 0.02--0.06, 
depending on the reaction type. A simple expression 
$\AN{} = {\FPT}(\PT{}){\GXA}(\XA{})$
can be used to approximate the experimental analyzing powers in the above
 range of high energies and $\PT{}$.
This scaling behaviour is better fulfilled 
for the $\pi^{+}$, $\pi^{0}$, $K^{+}$, $\eta$, and
$\Lambda$-production in $p^{\uparrow} \!p$-collisions, 
which takes place probably at the quark level.
The most solid experimental conformation of the $\XA{}$-scaling exists now
for $\pi^{+}$ production in  $p^{\uparrow} \!p(A)$-collisions, where 6
 independent measurements have been performed in a wide range of $\PT{}$,
$\XA{}$, and $\sqrt{s}$.
  
Significant non-asymptotic (energy dependent) contributions are observed
 for the $\pi^{-}$  and proton production. The former has a noticeable
 gluon contribution, and  the latter can be produced  mainly from protons,
 existing in the initial  state.

 The analyzing power for some reactions has not yet been  explored 
thoroughly enough to make a conclusion about the $\XA{}$-scaling features. 
The additional $\AN{}$-measurements are necessary at several c.m. angles 
in the central and fragmentation regions and at different energies. The bin 
size in  $\XA{}$ and $\PT{}$ should be small enough to get one unbiased 
averaging  over it, and to estimate mean values of $\XA{}$ and $\PT{}$ for 
each data point. In an ideal case, new experiments should measure
 $\XA{}$-dependence  at fixed $\PT{}$ and $\PT{}$-dependence at 
fixed $\XA{}$. Of interest is also a high $\PT{}$-region 
($2 \le \PT{} \le 10$ GeV/c), where the decrease of the  analyzing power is 
expected with a  $\PT{}$ rise according to some models \cite{RYSK,KANE,ANSE}.

 The asymptotic dependence of $\AN{}$ on $\XA{}$ for most of the hadrons has a
characteristic point $x_{0}$, where it intersects zero and probably changes
 its  sign. Such behaviour is in a qualitative agreement with the predictions
 from the models which take into account the Thomas precession and
 chromomagnetic forces between an outgoing quark and the rest of hadronic
 system. The linear dependence of $\AN{}$ on $\XA{}$ for most 
of the reactions may indicate that the polarization of a valence
quark, which is kicked out from a proton and fragments into a hadron $h$,
 containing this quark, is proportional to $\XA{}$ or to the secondary hadron
 energy.

  The use of eqs.~(\ref{eq:AN1}) -- (\ref{eq:F1p}) with the known parameters 
allows one to predict $\AN{}$ in a wide range of kinematic variables and to
use these predictions for the comparison with the models, to optimize future 
experiments  and to use some reactions as polarimeters.

\newpage
\listoffigures
\newpage
\clearpage
\clearpage
\clearpage
\clearpage
\clearpage
\clearpage
\clearpage
\clearpage
\clearpage
\clearpage
\clearpage
\clearpage
\clearpage
\clearpage
\clearpage
\clearpage
\clearpage
\clearpage
\clearpage
\clearpage
\clearpage
\clearpage
\clearpage
\clearpage

\end{document}